\documentclass[aps,pre,twocolumn,superscriptaddress,showpacs]{revtex4-1}
\usepackage{amsmath}
\usepackage{amssymb}
\usepackage{tikz}
\usepackage{graphicx}
\usepackage{dcolumn}
\usepackage{bm}
\usepackage{epsfig}
\usepackage{psfrag}
\def\II{\hbox{$1\hskip -1.2pt\vrule depth 0pt height 1.6ex width 0.7pt\vrule depth 0pt height 0.3pt width 0.12em$}}
\newcommand{\reffig}[1]{\mbox{Fig.~\ref{#1}}}
\newcommand{\refeq}[1]{\mbox{Eq.~(\ref{#1})}}
\newcommand{\refsec}[1]{\mbox{Sec.~\ref{#1}}}

\newcommand{\T}{${\mathcal T}\,$}
\newcommand{\Ti}{${\mathcal T}$}
\newcommand{\be}{\begin{equation}}
\newcommand{\ee}{\end{equation}}
\newcommand{\bal}{\begin{align}}
\newcommand{\eal}{\end{align}}
\newcommand{\ba}{\begin{eqnarray}}
\newcommand{\ea}{\end{eqnarray}}

\def\II{\hbox{$1\hskip -1.2pt\vrule depth 0pt height 1.6ex width 0.7pt\vrule depth 0pt height 0.3pt width 0.12em$}}

\begin{document}

\title{\bf Closed and open superconducting microwave waveguide networks as a model for quantum graphs}
\author{Barbara Dietz}
\email{barbara@ibs.re.kr}
\affiliation{%
Center for Theoretical Physics of Complex Systems, Institute for Basic Science (IBS), Daejeon 34126, Korea
}
\affiliation{%
Basic Science Program, Korea University of Science and Technology (UST), Daejeon 34113, Republic of Korea
}
\author{Tobias Klaus}
\affiliation{%
Institut f\"ur Kernphysik,Technische Universit\"at Darmstadt,D-64289 Darmstadt, Germany
}
\author{Marco Masi}
\affiliation{%
Institut f\"ur Kernphysik,Technische Universit\"at Darmstadt,D-64289 Darmstadt, Germany
}
\author{Maksym Miski-Oglu}
\email{M.MiskiOglu@gsi.de}
\affiliation{%
GSI Helmholtzzentrum f\"ur Schwerionenforschung GmbH
D-64291 Darmstadt, Germany}
\author{Achim Richter}
\email{richter@ikp.tu-darmstadt.de}
\affiliation{%
Institut f\"ur Kernphysik,Technische Universit\"at Darmstadt,D-64289 Darmstadt, Germany
}
\author{Tatjana Skipa}
\affiliation{%
Darmstadt University of Applied Sciences, Darmstadt, Germany.
}
\author{Marcus Wunderle}
\affiliation{%
Institut f\"ur Kernphysik,Technische Universit\"at Darmstadt,D-64289 Darmstadt, Germany
}

\date{\today}

\bigskip

\begin{abstract}
	We report on high-precision measurements that were performed with superconducting waveguide networks with the geometry of a tetrahedral and a honeycomb graph. They consist of junctions of valency three that connect straight rectangular waveguides of incommensurable lengths. The experiments were performed in the frequency range of a single transversal mode, where the associated Helmholtz equation is effectively one dimensional and waveguide networks may serve as models of quantum graphs with the joints and waveguides corresponding to the vertices and bonds. The tetrahedral network comprises T junctions, while the honeycomb network exclusively consists of Y junctions, that join waveguides with relative angles $90^\circ$ and $120^\circ$, respectively. We demonstrate that the vertex scattering matrix, which describes the propagation of the modes through the junctions strongly depends on frequency and is non-symmetric at a T junction and thus differs from that of a quantum graph with Neumann boundary conditions at the vertices. On the contrary, at a Y junction, similarity can be achieved in a certain frequeny range. We investigate the spectral properties of closed waveguide networks and fluctuation properties of the scattering matrix of open ones and find good agreement with random matrix theory predictions for the honeycomb waveguide graph. 
\end{abstract}

\bigskip
\maketitle

\section{Introduction\label{Intro}} 

Quantum graphs have been proposed three decades ago in Refs.~\cite{Kottos1997,Kottos1999} as a powerful testbed for the study of one central aspect of quantum chaos, namely the identification of signatures of classical chaos in the spectral properties of the corresponding quantum system~\cite{LesHouches1989,Guhr1998,Haake2018}. According to the Bohigas-Giannoni-Schmit conjecture (BGS)~\cite{Bohigas1984} these are described by the Gaussian ensembles of random matrix theory (RMT)~\cite{Mehta1990} of corresponding universality class. For a typical quantum system with a chaotic classical counterpart and preserved time-reversal (\Ti) invariance the appropriate ensemble is the Gaussian orthogonal ensemble (GOE) of real symmetric random matrices~\cite{Berry1979,Casati1980}. This central conjecture of quantum chaos was confirmed experimentally, e.g., with flat, cylindrical microwave resonators~\cite{Stoeckmann1990,Sridhar1991,Graef1992,Deus1995,StoeckmannBuch2000}. Here, the fact is exploited that below the cutoff frequency $f^{cut}$ of the first transverse-electric mode the electric-field strength is perpendicular to the top and bottom plane of the cavity, so that the associated Helmholtz equation is scalar and obeys Dirichlet boundary conditions (BCs) along the side wall. Thus it is mathematically identical to the Schr\"odinger equation of a quantum billiard of corresponding shape with these BCs. Accordingly, in that frequency range such cavities are referred to as microwave billiards. Complete sequences of up to 5000 eigenfrequencies~\cite{Richter1999,Dietz2015a,Dietz2019a} were obtained in high-precision measurements at liquid-helium temperature $T_{\rm LHe}=4.2$~K with niobium and lead-coated microwave billiards which become superconducting at $T_c=9.2$~K and $T_c=7.2$~K, respectively. Furthermore, the fluctuation properties of the scattering ($S$) matrix of open quantum systems with preserved and also with partially violated \T invariance where studied in experiments at room temperature and analyzed with regard to exact analytical results that were derived based on the Heidelberg approach for quantum-chaotic scattering~\cite{Verbaarschot1985,Dietz2007,Dietz2009,Dietz2010}. 

Quantum graphs, consisting of bonds that are connected at vertices~\cite{Kottos1997,Kottos1999,Pakonski2001,Texier2001,Kuchment2004,Gnutzmann2006,Berkolaiko2013}, were introduced in~\cite{Pauling1936} and have found applications in context with quantum wires~\cite{Sanchez1998,Kostrykin1999}, optical waveguides~\cite{Mittra1971}, mesoscopic quantum systems~\cite{Kowal1990,Imry1996}, and waveguide networks~\cite{Post2012,Exner2015,Gnutzmann2022,Zhang2022,Ma2022}. They are governed by the one-dimensional Schr\"odinger equation along the bonds with BCs at the vertices, that impose continuity upon the wave functions passing through them and current conservation. Depending on the BC the fluctuation properties in the eigenvalue spectra of closed quantum graphs with incommensurable bond lengths coincide with those of random matrices from the GOE~\cite{Gnutzmann2004}. The BCs can be expressed in terms of unitary vertex matrices~\cite{Kottos1999,Kostrykin1999,Gnutzmann2006,Harrison2007,Lawniczak2019} defining the propagation of the waves through the vertices or scattering off them. Actually, the ergodicity of the wave dynamics of such quantum graphs~\cite{Roth1983,Keating1991,Kottos1999} originates from the scattering characteristics of the waves entering and exiting a vertex~\cite{Gnutzmann2006}. 

The $S$-matrix two-point correlation functions associated with the scattering dynamics of open quantum graphs with incommensurable bond lengths, that are coupled to their environment through bonds extending to infinity, were shown to coincide with those of random $S$ matrices deduced from the Heidelberg approach for quantum-chaotic scattering~\cite{Verbaarschot1985,Pluhar2013,Pluhar2013a,Pluhar2014,Fyodorov2005}. Thus, even though closed and open quantum graphs are governed by the one-dimensional Schr\"odinger equation, their wave dynamics may exhibit the same features as typical quantum systems with a chaotic classical dynamics. Another advantage of quantum graphs is, that all three universality classes associated with Dyson's threefold way~\cite{Dyson1962,Joyner2014} can be simulated for $\delta$-type boundary conditions at the vertices~\cite{Kottos1997,Kottos1999,Pakonski2001,Kuchment2004}. 

Quantum graphs with wave-chaotic dynamics subject to Neumann BCs at the vertices, named 'Neumann quantum graphs' in the following, have been simulated with microwave cable networks~\cite{Hul2004}. These consist of coaxial cables that are coupled by either commercial T joints or by joints with higher connectivity that are designed such that they fulfill these BCs and have been realized for all three Wigner-Dyson universality classes~\cite{Lawniczak2010,Bialous2016,Rehemanjiang2016,Martinez2018,Martinez2019,Lu2020,Che2021}, also for the universality classes of the ten-fold way~\cite{Rehemanjiang2020,Altland1997}. Properties of the corresponding open quantum graphs with wave chaotic dynamics have been investigated experimentally in~\cite{Lawniczak2008,Lawniczak2011,Hul2012,Lawniczak2014,Hul2012,Lawniczak2020,Rehemanjiang2016,Martinez2018,Martinez2019,Lu2020,Chen2020,Chen2021,Bialous2023}. A drawback of Neumann quantum graphs, and therefore also of microwave cable networks, is the occurrence of backscattering at their vertices entailing modes that are localized on single bonds or on a fraction of the bonds and therefore do not exhibit the complexity of typical quantum systems with chaotic classical counterpart. These modes are non-universal in the sense that their formation depends on the lengths of the involved bonds, yet they can be prevented by an appropriate choice of the BCs. Furthermore, wave functions are experimentally accessible at the joints of a microwave cable network, but not along the bonds.

Complete sequences of eigenfrequencies of microwave billiards may be determined with high precision only in superconducting resonators. However, even microwave cable networks made from niobium are not superconducting at $T_{\rm LHe}$, because they are filled with a dielectric medium. Accordingly, we proposed in 2014 microwave waveguide networks as an alternative microwave setup for the modeling of properties of quantum graphs and report in the present work on the high-precision measurements that were performed with superconducting cavities at $T_{\rm LHe}$ during 2014-2015. They consist of straight waveguides, whose widths are larger than their height, with Dirichlet BCs at the walls, that are connected at junctions. Such systems may serve as models for quantum graphs in the frequency range of a single transversal mode~\cite{Post2012,Exner2015,Gnutzmann2022,Zhang2022,Ma2022}, where waves propagating through the waveguides are governed by the one-dimensional Helmholtz equation and, accordingly, are referred to as waveguide graphs in the following. Note that, in distinction to microwave cable networks and Neumann quantum graphs, the vertex scattering matrix describing the propagation of the waves through the junctions, which can be obtained by proceeding as in~\cite{Bittner2013}, where it was computed analytically for bent waveguides, depends on the wavenumber. In~\cite{Zhang2022} wave functions were measured in a large-scale honeycomb waveguide network, that was designed based on the results presented in this article, at room temperature,  confirming predictions on the properties of those of the corresponding quantum graph~\cite{Kaplan2001}. In the experiments at $T_{\rm LHe}$ properties of the wave functions may be studied in terms of the strength distribution~\cite{Dembowski2005}. 

The present work is organized as follows. In \refsec{MWN} we briefly introduce quantum graphs, microwave cable networks and then waveguide graphs and their design. They consist of waveguides that are joined by junctions of connectivity three. We performed numerical simulations with CST Studios, in which three waveguides were joined at varying relative angles. The objective was to find an optimal design for the waveguide graphs, in which non-universal contributions are minimized and spectral properties are as close as possible to RMT predictions. We show results for T and Y junctions, that join three waveguides with relative angles $90^\circ$ and $120^\circ$, respectively. In~\refsec{Exp} we describe the experimental setup which we used for the measurements at superconducting conditions.
\begin{figure}[!th]
\includegraphics[width=0.7\linewidth,height=3cm]{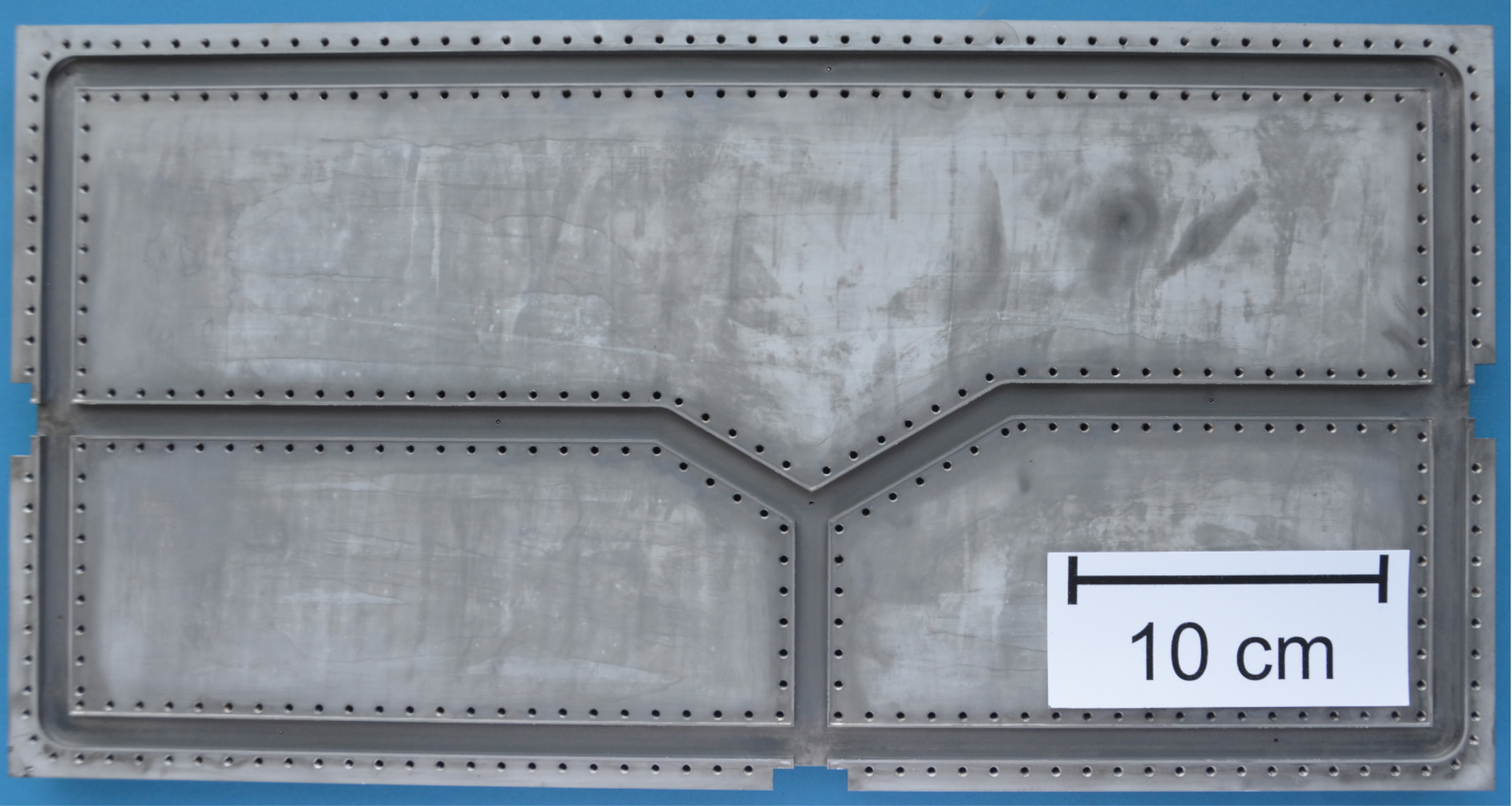}
\includegraphics[width=0.7\linewidth,height=3cm]{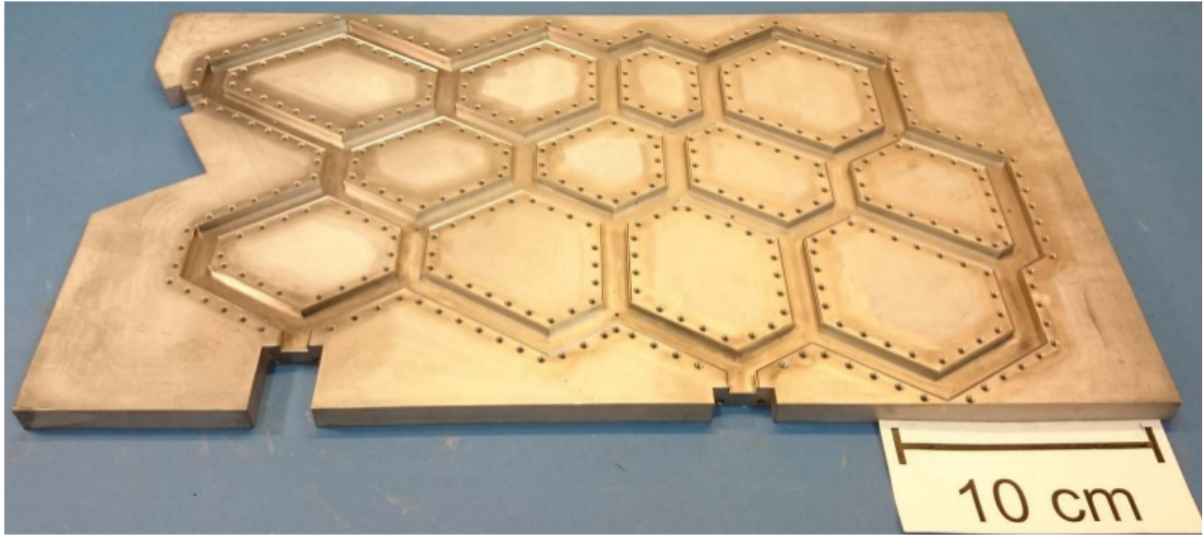}
\caption{Top: Bottom plate of the tetrahedral waveguide graph, exhibiting the channel system, the screw holes and the three cutouts, where the waveguide-to-coaxial adapters are attached. Bottom: Same as top for the honeycomb waveguide graph.}
\label{fig:photos}
\end{figure}
We considered two waveguide graphs whose geometries are shown in~\reffig{fig:photos}. The waveguide graph exhibited in the top panel has 4 vertices, namely three T junctions at the left, right and bottom side, respectively, and one Y junction at the center, and 6 bonds, and thus mimicks a tetrahedral graph, so we use this denomination when referring to it in the following. Furthermore, it comprises four $90^\circ$ bends. We used this design in the first experiment, because it provides the simplest realization of a quantum-chaotic graph and has been used in numerous studies, e.g. in Refs.~\cite{Kottos1999,Hul2004,Bittner2013,Gnutzmann2013,Chen2020}. However, it is well known~\cite{Kottos1999,Dietz2015c} that, in order to realize a quantum graph whose short \emph{and} long range spectral properties agree well with RMT predictions, the number of vertices and connectivity should be chosen as large as possible, and the geometry of the junctions should be chosen such that backscattering is minimized~\cite{Dietz2015c,Dietz2017b}.  The size of the microwave cavity is restricted by that of the cryostat used for cooling to liquid-Helium temperature (see below), and thus, the number of vertices and bends is limited by the size of the waveguides. Accordingly, we performed numerical simulations with CST Studios to identify an optimal geometry, and came to the conclusion that, preferably all corners and relative angles of waveguides at junctions should be chosen $\gtrsim 120^\circ$. These simulations led to the waveguide graph shown in the bottom panel of~\reffig{fig:photos}, which consists of 22 Y junctions and 16 $120^\circ$ bends and is named honeycomb graph in the following because its structure resembles that of a honeycomb. Then, in~\refsec{Spectral} we review our results on the spectral properties of waveguide graphs. Deviations from RMT predictions may be due to the geometry of the waveguide networks, like a too low number of vertices, which would also become visible in those of the corresponding quantum graph. Or it may originate from a lack of complexity of the wave dynamics and non-universal effects resulting from the particular scattering features at their junctions and bends. Therefore, we compare the spectral properties with those of the corresponding Neumann quantum graph, which may exhibit such drawbacks due to the BCs at the vertices, and with RMT predictions. In~\refsec{Smatrix} we summarize the results for the strength distribution and fluctuation properties of the $S$ matrix associated with the measurement process employed to obtain resonance spectra. The scattering experiments were performed at superconducting conditions and calibration was performed with  the so called \textbf{T}hru--\textbf{R}eflection--\textbf{L}ine (TRL) method introduced in~\cite{Marks1991,Rytting2001}. Microwaves were coupled into the cavity through openings located at three vertices or bends via waveguide-to-coaxial adapters. Due to superconducivity absorption into the walls is negligible, so that scattering takes place exclusively via these openings. In~\refsec{Concl} we summarize our findings and discuss them.   

\section{Quantum Graphs, Microwave Cable Networks, and Microwave Waveguide Networks\label{MWN}} 
\subsection{Quantum graphs\label{QG}}
A quantum graph comprises $\mathcal{V}$ vertices $i=1,\dots,\mathcal{V}$ connected by $\mathcal{B}$ bonds. Its geometry is defined by the connectivity matrix $\hat C$ with diagonal elements $C_{ii}=0$ and off-diagonal elements $C_{ij}=1$ if vertices $i$ and $j$ are connected by bonds, whose lengths are denoted by $L_{ij}$, and $C_{ij}=0$ otherwise. The associated wave-function components $\psi_{ij}(x)$ are governed by the one-dimensional Schr\"odinger equation
\be
\frac{d^2}{dx^2}\psi_{ij}(x)+k^2\psi_{ij}(x)=0,\label{WE}
\ee
where $k$ denotes the wavenumber and the coordinate $x$ increases from $x=0$ at vertex $i$ to $x=L_{ij}$ at vertex $j$. They are subject to BCs imposed at vertices $i$ and $j$ that ensure continuity,
\begin{equation}
\psi_{ij}(x=0)=\varphi_i,\, \psi_{ij}(x=L_{ij})=\varphi_j,\, i<j,
\label{BC1}
\end{equation}
and current conservation. For Kirchhoff BCs, which is a special case of $\delta$-type BCs~\cite{Kottos1999,Kostrykin1999,Texier2001,Kuchment2004,Gnutzmann2006,Berkolaiko2013}, they are given by
\begin{equation}
-\sum_{j<i}C_{ji}\frac{d}{dx}\psi_{ji}(x=L_{ij})+\sum_{j>i}C_{ij}\frac{d}{dx}\psi_{ij}(x=0)=\lambda_i\varphi_i.
\label{BC2}
\end{equation}
At each vertex $i$ a unitary matrix $\hat\sigma^{(i)}_{ji,im}$ with dimension given by its valency $v_i$, which depends on the BCs, determines the propagation of a wave function passing through it from vertex $m$ to vertex $j$,
\ba
&&\sigma^{(i)}_{ji,im}=\left(-\delta_{j,m}+\frac{1}{v_i}\left[1+\frac{\left(1-i\Lambda_i\right)^2}{1+\Lambda_i^2}\right]\right)C_{ij}C_{im},\label{Sigma}\\
&&\Lambda_i=\frac{\lambda_i}{v_ik}.\nonumber
\ea
For Dirichlet BCs $\lambda_i\to\infty$, corresponding to completely decoupled bonds, it becomes $\sigma^{(i)}_{ji,im}=-\delta_{j,m}C_{ij}C_{im}$, whereas Neumann BCs, $\lambda_i=0$,  yield
\be
\sigma^{(i)}_{ji,im}=\left(-\delta_{j,m}+\frac{2}{v_i}\right)C_{ij}C_{im}.\label{SigmaN}
\ee
In between these limits it depends on $k$. The eigenwavenumbers $k_n=\frac{2\pi f_n}{c}$, with $f_n$ denoting the corresponding eigenfrequencies, of a quantum graph are determined numerically by varying the wavenumber $k$ and searching for the non-trivial solutions of the set of homogenoues equations $\hat h(k)\boldsymbol{\varphi}=\boldsymbol{0}$ resulting from the BCs, that is, of
\begin{equation}
\label{QuantCond}
\det\hat h(k)=0.
\end{equation}
Here the components of $\boldsymbol{\varphi}$ correspond to the wave function components $\varphi_i$ and the matrix elements of $\hat h$ are given by~\cite{Kottos1999}
\begin{equation}
h_{ij}(k)=\left\{{\begin{array}{cc}
        -\sum_{m\ne i}\cos\left(kL_{im}\right)\frac{C_{im}}{\sin\left(kL_{im}\right)}-\frac{\lambda_i}{k}&, i=j\\
        \frac{C_{ij}}{\sin(kL_{ij})}&, i\ne j\\
        \end{array}}
        \right.\, .\label{QuantE}
\end{equation}
The average spectral density is constant and given by Weyl's law,
\be
\label{Weyl}
\langle\rho(f_n)\rangle=\frac{2\mathcal{L}}{c},\, \langle\rho(k_n)\rangle =\frac{\mathcal{L}}{\pi},
\ee
where $\mathcal{L}$ denotes the total length of the graph, $c$ is the speed of light in vacuum, and $\langle\cdot\rangle$ means spectral averaging and may include an ensemble average over several measurements. The components of the associated eigenvectors yield the values of the wave functions at the vertices $i$, $\varphi_i$, and thus the eigenfunctions~\cite{Kottos1999}. 

The closed quantum graph is converted into a scattering system with $M$ scattering channels by attaching to $M$ vertices leads that extend to infinity. The fluctuation properties of the associated $S$ matrix have been shown to comply with those of quantum-chaotic scattering processes~\cite{Pluhar2013,Pluhar2013a,Pluhar2014}. Actually, the $M\times M$ $S$ matrix describing the scattering process of the waves entering the quantum graph through one of these leads from infinity and exiting it through the same (reflection) or another lead (transmission) can be brought to the form
\be
\hat S_{\mathcal{V}}(k)=\II -2\pi i\hat W^T\left(\hat h(k)+i\pi\hat W\hat W^T\right)^{-1}\hat W.
\label{SVertex}
\ee
It is similar to that derived on the basis of the $S$-matrix formalism for compound nucleus reactions~\cite{Mahaux1969} except for the structure of $\hat h(k)$, which depends linearly on $k$ in the RMT model. 
\subsection{Microwave cable networks\label{MN}}
Below the cutoff frequency for the first transverse electric (TE$_{11}$) mode only the fundamental transverse electromagnetic modes (TEM) can propagate between the conductors~\cite{Jones1964,Savytskyy2001}. These are governed by the one-dimensional telegraph equation, which obeys the continuity equation and for conventional joints~\refeq{BC2} with $\lambda_i=0$, and thus is mathematically identical to the Schr\"odinger equation~\refeq{WE} of a quantum graph with Neumann BCs at the vertices and corresponding geometry~\cite{Hul2004}. The analogy between a Neumann quantum graph and a microwave cable network of corresponding geometry is exact for lossless coaxial cables, that is, for vanishing Ohmic resistance. Generally, the determination of complete sequences of eigenfrequencies is involved due to the unavoidable Ohmic losses and was achieved only in experiments with parametric quantum graphs~\cite{Rehemanjiang2016,Lu2020}.

\subsection{Microwave waveguide networks\label{WG}}
\begin{figure}[htbp]
\includegraphics[width=0.6\linewidth]{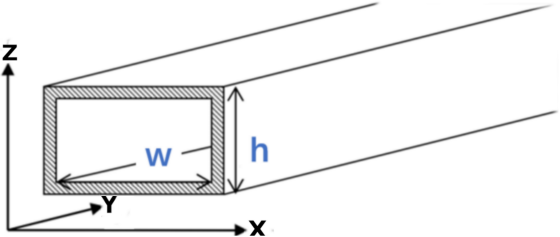}
	\caption{Schmematic of a rectangular waveguide.}
\label{WGScheme}
\end{figure}

A waveguide network is constructed from metallic rectangular waveguides of width $w$ and height $h<w$ schematically shown in~\reffig{WGScheme}. 
Their lengths are incommensurable and, respectively three of them are connected. Below the cutoff frequency $f_{max}=c/2h$ of the first transverse-electric mode the electric field strength is perpendicular to the top and bottom walls of the waveguides, that is, in a coordinate system with $z$-axis in that direction and $x$ and $y$ axes in transversal and longitudinal directions, respectively, $\vec E =E_z\vec e_z$. In that frequency range the microwaves are governed by the two-dimensional Helmholtz equation for a perfect electric conductor, that is, with Dirichlet BCs at the side walls,
\be
\left[\frac{\partial^2}{\partial x^2}+\frac{\partial^2}{\partial y^2}+k^2\right]E_z=0,\, (x,y)\in\Omega,\, E_z\vert_{x,y\in\partial\Omega}=0.
\ee
Here, $k=\frac{2\pi f}{c}$ denotes the wavenumber. The wavenumbers in longitudinal direction, $k_{y;n}$, are determined from the ordered eigenfrequencies $f_n$, $f_1\leq f_2\leq\dots$ through the relation 
\be\label{kym}
k_{y;n}=\sqrt{\left(\frac{2\pi f_n}{c}\right)^2-\left(\frac{q\pi}{w}\right)^2-\left(\frac{p\pi}{h}\right)^2}
\ee
where the indices $q$ and $p$ count the number of modes excited in transversal and $z$ direction~\cite{Jackson1999}, respectively. We chose in the experiments and numerical simulations waveguides of width $w=10.668$~mm and height $h=4.318$~mm. Then, the cutoff frequencies $f_{\rm TM_{qp}}$ of the first and second transversal mode and for the first excited transverse-magnetic mode, ${\rm TM_{01}}$, are given as
\ba\label{Eq:fre}
	f_{\rm TM_{10}} &= \frac{c}{2w}\, &\approx 14\, {\rm GHz}, \\
	f_{\rm TM_{20}} &= \frac{c}{w}\, &\approx 28\, {\rm GHz},\\
	f_{\rm TM_{01}} &= \frac{c}{2h}\, &\approx 34\, {\rm GHz},
\ea
respectively.
In the frequency range $f_{\rm TM_{10}}\leq f\leq f_{\rm TM_{20}}$ a single transversal mode is excited, so that the wave equation is effectively one-dimensional, that is, coincides with that on the bonds of a quantum graph or microwave cable network. Accordingly, in that frequency range a waveguide network can be used to simulate properties of quantum graphs with BCs at the vertices that differ from Neumann and Dirichlet BCs. 

\begin{figure}[htbp]
\includegraphics[width=0.49\linewidth]{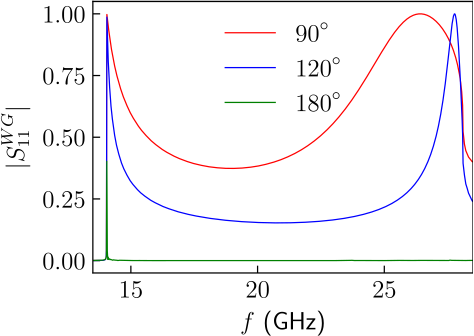}  
\includegraphics[width=0.49\linewidth]{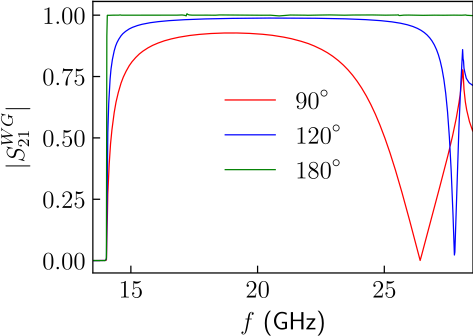}  
	\caption{Numerical simulation of the $S$ matrix properties of waveguides with bending angles $90^\circ$ (red) and $120^\circ$ (blue), and of a straight waveguide (green) corresponding to a bending angle $180^\circ$.}   
\label{SmatrWG}   
\end{figure}

It should be noted that waveguide bends correspond to vertices with valency two~\cite{Zhang2022}, whereas coupling two bonds of a quantum graph or two coaxial cables of a microwave cable network yields a bond whose length equals that of the sum of their lengths. This is illustrated in~\reffig{SmatrWG} where the $S$ matrix $\hat S^{WG}$ for the scattering through waveguides with bending angles $90^\circ$ (red) and $120^\circ$ (blue) is plotted. We used CST Microwave Studio to compute the $S$ matrix for these cases and also for T- and Y-junctions by solving the three-dimensional Maxwell equations imposing the BCs for perfectly conducting metallic walls. The straight waveguide parts had finite lengths and those not used as scattering channels were terminated with 50 Ohm ports. Note that the $S$ matrix is ill-defined at the cutoff frequencies were a new scattering channel opens~\cite{Jackson1999,Bittner2013}, leading to the kink observed, e.g. in~\reffig{SmatrWG} at $f\simeq 14$~GHz for $\vert S_{11}^{\rm WG}\vert$ and for $\vert S_{21}^{\rm WG}\vert$ at the cutoff frequency of the second transversal mode $f\simeq 28$~GHz. Detailed experimental and analytical results on the $S$-matrix properties of bent waveguides can be found in Ref.~\cite{Bittner2013}. In that work the occurrence of bound states in the region of the bend was investigated as function of the bending angle. It was attributed to scattering between the inner corner and the opposite sidewall and demonstrated to lead to a worsening of transmittance through the waveguides. These are unwanted features that can be avoided by choosing bending angles of at least $120^\circ$. 

\begin{figure}[htbp]
\includegraphics[width=0.49\linewidth]{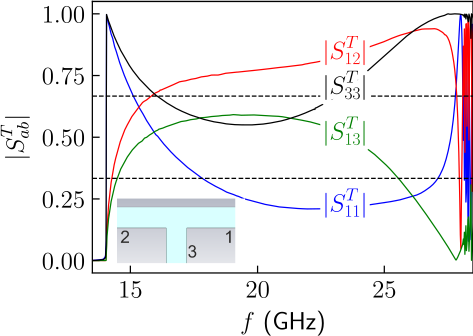}  
\includegraphics[width=0.49\linewidth]{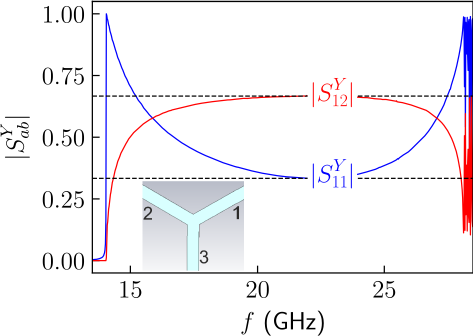}  
	\caption{Numerical simulation of the properties of the $S$ matrix $S^T$ of a T-junction (left) and $\hat S^Y$ of a Y-junction (right) joining three waveguides as illustrated in the insets. The dashed lines indicate the values of a vertex $S$ matrix with valency three and Neumann BCs~\refeq{NeumannS}.}   
\label{SmatrYT}   
\end{figure}

We compare the spectral properties of the waveguide graphs that we investigated experimentally to those of quantum graphs of same geometry. In distinction to the bends, valencies at the junctions of waveguide graphs coincide with those of the corresponding quantum graph. The valencies equaled three in all realizations. Furthermore, the vertex $S$ matrix depends on $k$ and on the bending angle, whereas for Neumann quantum graphs it is constant~\cite{Kottos1999,Hul2004},
\be
 S^{NQG}_{ab}=2/3-\delta_{ab}.
\label{NeumannS}
\ee
In~\reffig{SmatrYT} the $S$-matrix elements $\hat S^T$ and $\hat S^Y$ of a T- and a Y-junction are plotted as function of frequency $f$ for $f_{\rm TM_{10}}\lesssim f\lesssim f_{\rm TM_{20}}$. 
The $3\times 3$-dimensional $S$ matrix is symmetric in all cases, $S^{\alpha}_{ab}=S^{\alpha}_{ba}, a,b=1,2,3, \alpha =T,Y$. Yet, for the T junction $S^T_{13}=S^T_{23}$ and $S^T_{11}=S^T_{22}$, however, e.g., $S^T_{13}\ne S^T_{12}$, $S^T_{11}\ne S^T_{33}$, implying that $\hat S^T$ differs significantly from $\hat S^{NQG}$ given in~\refeq{NeumannS} in the whole frequency range~\footnote{Note, that these results differ from those obtained in~\cite{Ma2022}, where the waveguide network was constructed from a square lattice}. On the contrary, for the Y junction all the transmission spectra, respectively, reflection spectra coincide, and $\hat S^Y$ is similar to that for Neumann BCs~\refeq{NeumannS}, and thus to that at the T joints of the corresponding microwave cable network in the range from $\approx 20-25$~GHz. The deviations of $\hat S^T$ and $\hat S^Y$ from $\hat S^{NQG}$ in~\refeq{NeumannS} are attributed to backscattering from the inner corners, which is enhanced for the former as may be expected from the results on bent waveguides obtained in Ref.~\cite{Bittner2013}. They are considerably larger for the $90^\circ$ bend. On the other hand, transmission is close to that of a straight waveguide (green) for the $120^\circ$ bend in the frequency range from $\approx 16$~GHz to $\approx 25$~GHz.

\section{Experimental setup\label{Exp}}
\begin{figure}[htbp]
\includegraphics[width=0.8\linewidth]{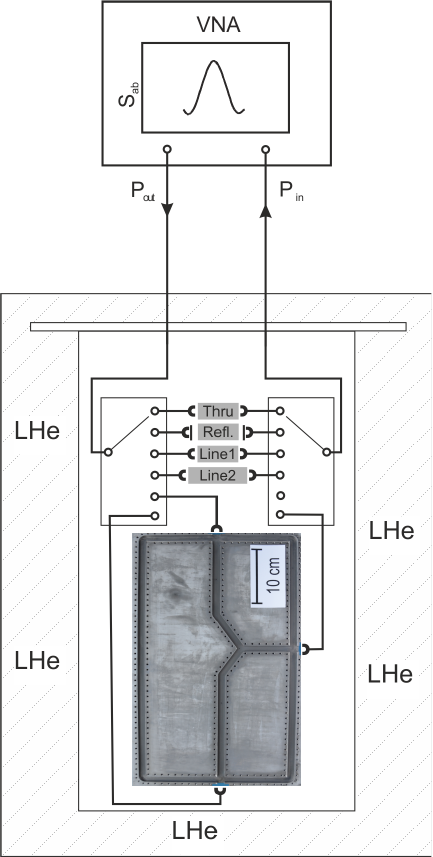}  
\caption{Schematic of the waveguide graph in the cryostat and the TRL setup. Microwaves are coupled into and out of the cavity with a VNA.}   
\label{fig:Cryostat}   
\end{figure}

The tetrahedral and honeycomb waveguide graphs were constructed from a top and bottom plate shown in~\reffig{fig:photos} The channels were milled out of a brass plate. Both the bottom and top plates were galvanically lead plated to achieve superconductivity at temperature $T_{\rm LHe}=4.2$~K. They were screwed together tightly through holes along the channels, recognizable in~\reffig{fig:photos}. Furthermore, grooves with width 1.3~mm and depth 1~mm were milled out of the bottom plate along the channels. Into these lead wire was inserted to ensure a good electrical contact between the top and bottom plates along all waveguides. The graphs were cooled down to 4.2~K in a cryostat that was filled with liquid Helium as illustrated for the tetrahedral graph in~\reffig{fig:Cryostat}. For this they were placed together with rf-switches and, for the scattering experiments, the calibration tool, into a cylindrical copper chamber that was evacuated to vacuum.  
\begin{figure}[htbp]
\includegraphics[width=\linewidth]{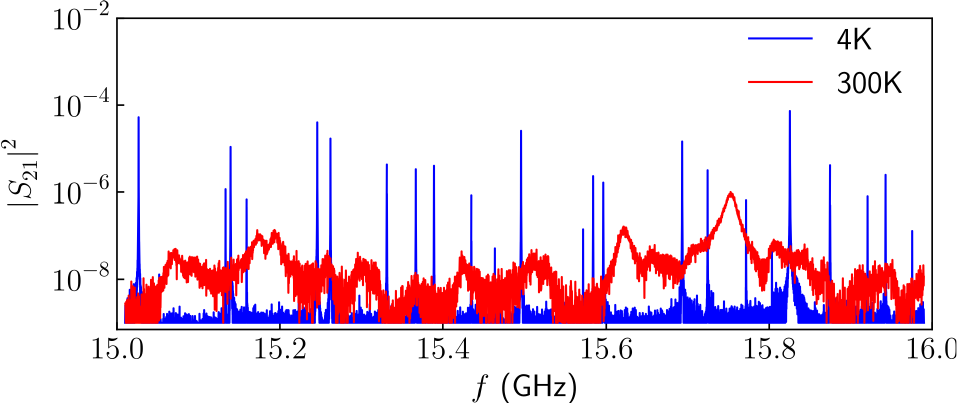}  
	\caption{Comparison of a part of a transmission spectrum, that was measured with the honeycomb waveguide graph at 300K (red line) with the corresponding one obtained at 4K (blue line).}   
\label{fig:SpectraComp}   
\end{figure}

We performed two types of experiments for each waveguide graph, namely one to determine the eigenfrequencies of the \emph{closed} graph, and a second one to measure the $S$ matrix of the corresponding \emph{open} graph. In the first one resonance spectra were measured with eight tiny wire antennas, that were attached to the top plate to couple the resonator modes as weakly as possible to the exterior so that all resonances are excited while at the same time a resonator with high-quality factor is realized which is optimally closed. Actually, this procedure is identical to that which was employed in our group in the experiments with superconducting microwave billiards~\cite{Dietz2015} and it was demonstrated explicitely, e.g. in~\cite{Graef1992,Dietz2014}, that no calibration of the spectra is required in these measurements to determine  with high precision the eigenvalues of the corresponding quantum billiard. 

For the scattering experiments the waveguide graph was opened at three junctions or bends visible in~\reffig{fig:photos}, and waveguide-to-coaxial adapters were tightly attached~\cite{Bittner2013,Zhang2022}, referred to as ports in the following. The objective was to study properties of the $S$ matrix describing the scattering process that exclusively takes place between the three ports. This is possible in experiments with superconducing resonators, where absorption into the walls is negligible, so that the ports may be considered to a good approximation as the only scattering channels.  Note, that for the determination of the eigenfrequencies in the measurements with antennas the port openings were closed with lead-plated metallic pieces whose shapes were designed such that this yielded a waveguide with straight wall in the tetrahedral graph and a $120^\circ$ bend in the honeycomb graph. Furthermore, the width $w$ and height $h$ of the waveguides was fixed by the impedance-matching condition with commercially available adapters to ensure a reflectionless escape of microwaves through the ports. 

Microwaves are sent from and received at a \textbf{V}ectorial \textbf{N}etwork \textbf{A}nalyser (VNA, model PNA N5230A by Agilent Technologies) through coaxial cables that are connected to the cavity via commercially available vacuum feedthroughs of SMA type. They are attached to the waveguide network at two antennas or ports denoted $a$ and $b$. The VNA measures relative phases $\phi_{ba}$ and ratios of the microwave power of the outcoming and ingoing microwaves, $\frac{P_{out,b}}{P_{in,a}}=|S_{ba}|^2$, that is, the complex scattering matrix elements $S_{ba}=|S_{ba}|e^{i\phi_{ba}}$. It describes the scattering process encountered by the waves inside the waveguide network when traveling from antenna or port $a$ to antenna or port $b$. In order to avoid cooling down for each antenna or port combination, the antenna or port cables were attached to a switch, which was inserted together with the waveguide graph into the copper box, and connected, respectively, two of them with the VNA. The advantage of superconducting microwave billiards is demonstrated in~\reffig{fig:SpectraComp}. The red spectrum was measured at room temperature, the blue one at superconducting conditions. The drastic reduction of Ohmic losses gives rise to the very narrow resonances, a prerequisite to determine a complete sequence of eigenfrequencies, that are obtained from the positions of the resonances. 

The $S$ matrix governing the measurement process of a resonator coupled to a VNA via $M$ open channels that support respectively one open channel is given by~\cite{Albeverio1996} 
\begin{equation}\label{eq:SResonator}
	\hat S(f) = \II - i\hat{W}^\dagger\left(f\II-\hat{H}^{Res}+\frac{i}{2}\hat W\hat W^\dagger\right)^{-1}\hat{W}.
\end{equation}
Here, $\hat H^{Res}$ denotes the resonator Hamiltonian and $\hat W$ describes the coupling of the resonator modes to the $M$ channels. Close to an isolated resonance located at the eigenfrequency $f_n$, $|S_{ba}|$ is well described by the complex Breit-Wigner form,
\begin{equation} \label{Eq:bw}
	S_{ba}(f)=\delta_{ba}-i\frac{\sqrt{\gamma_{na}\gamma_{nb}}}{f-f_n+\frac{i}{2}\Gamma_n},
\end{equation}
with $\gamma_{na}$ and $\gamma_{nb}$ denoting the partial widths associated with antennas $a$ and $b$ and $\Gamma_n$ the total width, which is given by the sum of the partial widths and the width $\Gamma_{abs}$ due to absorption in the walls of the waveguide. The individual partial widths, that are proportional to the modulus of the wave functions at the positions of the antennas, could only be determined in experiments that were performed at 2~K in~\cite{Alt1995}, however, information on the wave function components can also be obtained from the resonance strengths~\cite{Dembowski2005,Dietz2006}. The resonance parameters, that is, the resonance strengths $\gamma_{na}\gamma_{nb}$, resonance widths $\Gamma_n$ and eigenfrequencies $f_n$ are determined by fitting the complex Breit-Wigner form~\refeq{Eq:bw} to the resonances in the transmission ($a\ne b$) and reflection ($a=b$) spectra~\cite{Dembowski2005}. This is feasible in measurements at superconducting conditions because there the widths of the resonances are small compared to the average spacing between them. Thereby, a complete sequence of 282 eigenfrequencies was obtained from the uncalibrated spectra measured with antennas, and the resonance parameters of 187 resonances could be determined from the calibrated spectra measured with the three ports. In the latter case part of the resonaces were weakly overlapping. This is attributed to the larger openness~\cite{Bialous2020} of the resonator compared to that with short antennas, that is, the stronger coupling of resonator modes to the exterior through the ports.

To obtain information on the properties of the $S$ matrix of open waveguide graphs we measured the $S$ matrix using the ports. However, impedance mismatches at connectors and attenuation in the coaxial cables are unavoidable and lead to systematic errors in the measurement of the $S$ matrix. To minimize them an appropriate calibration procedure is required which is applicable to measurements at superconducting conditions. For this we employed the so called \textbf{T}hru--\textbf{R}eflection--\textbf{L}ine (TRL) calibration method introduced in~\cite{Marks1991,Rytting2001} which is widely used for measurements under cryogenic conditions~\cite{Laskar1996}. In this method, illustrated in~\reffig{fig:Cryostat}, the two coaxial cables exiting from the VNA are connected to six-position microwave switches that couple them either to the cavity, a Thru standard, a Reflect standard consisting of two identical reflectors attached to the ends of coaxial cables, or two Line standards. For the latter we used two coaxial cables with different lengths and both shorter than the typical wave length in the experiment. Actually, the TRL procedure fails when the frequency hits a resonance frequency of one of these cables, so that the choice of their lengths is crucial. We chose them such, that we could analyze properties of the $S$ matrix in the ranges $[14.5,19.5]$~GHz and  $[21,26.5]$~GHz that are in the region of a single transversal mode. 

Without the switches the measurement of the $S$ matrix of the different standards would require up to four cool-down and warm-up cycles. Yet, the use of latching microwave switches~\cite{Ranzani2012,Yeh2013} allows the measurements for all calibration standards and the microwave waveguide network within one thermal cycle. We employed two cryogenic 6-position switches (Radiall coaxial subminiature latching switches R591722605) for the measurement of the $S$ parameters of the Thru, Reflect, the two Line standards and of the microwave waveguide network. The final calibration of the $S$ matrix was done offline with the procedure described in detail in Ref.~\cite{Rytting2001}. 

\begin{figure}[htbp]
\includegraphics[width=0.49\linewidth]{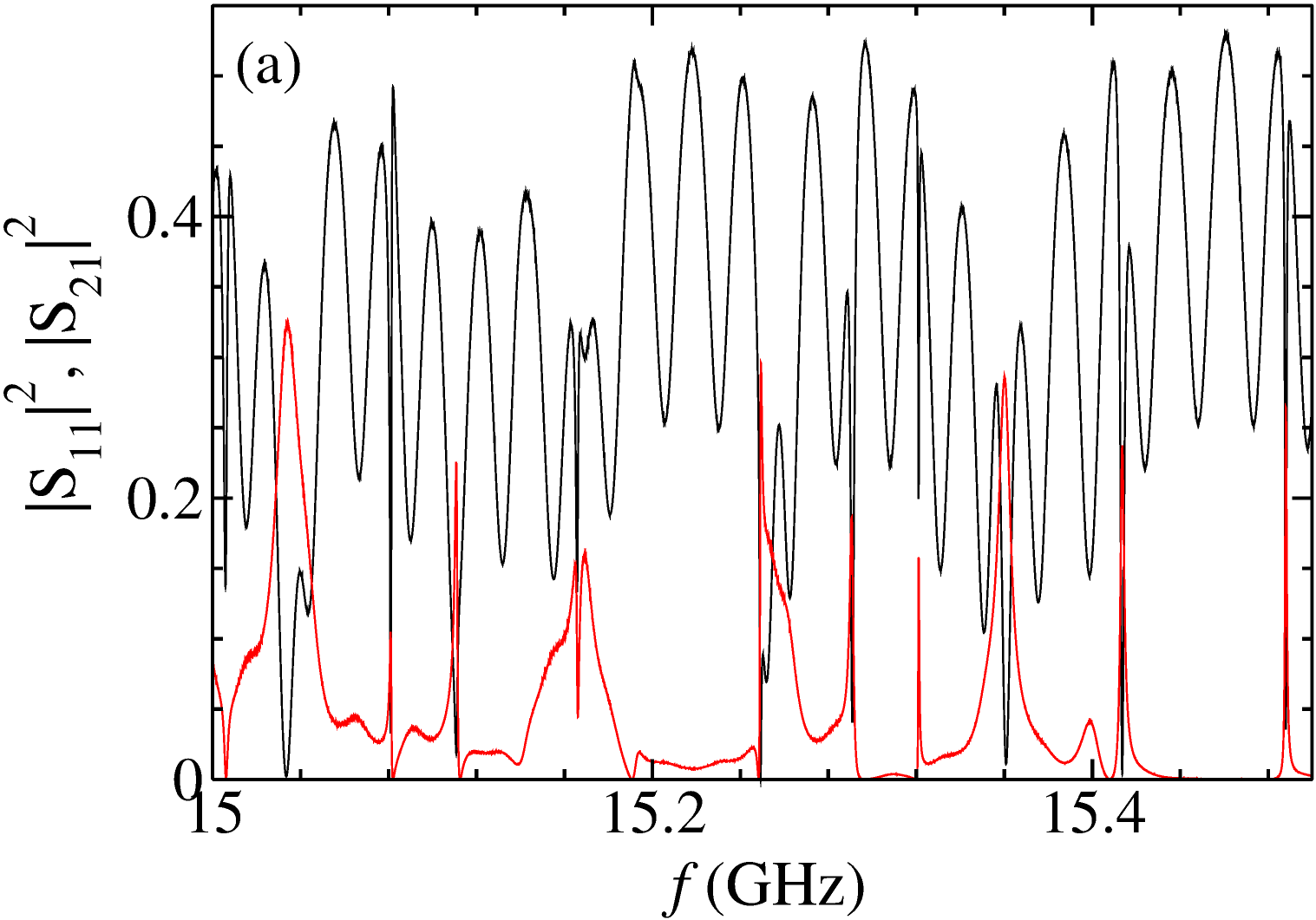}  
\includegraphics[width=0.49\linewidth]{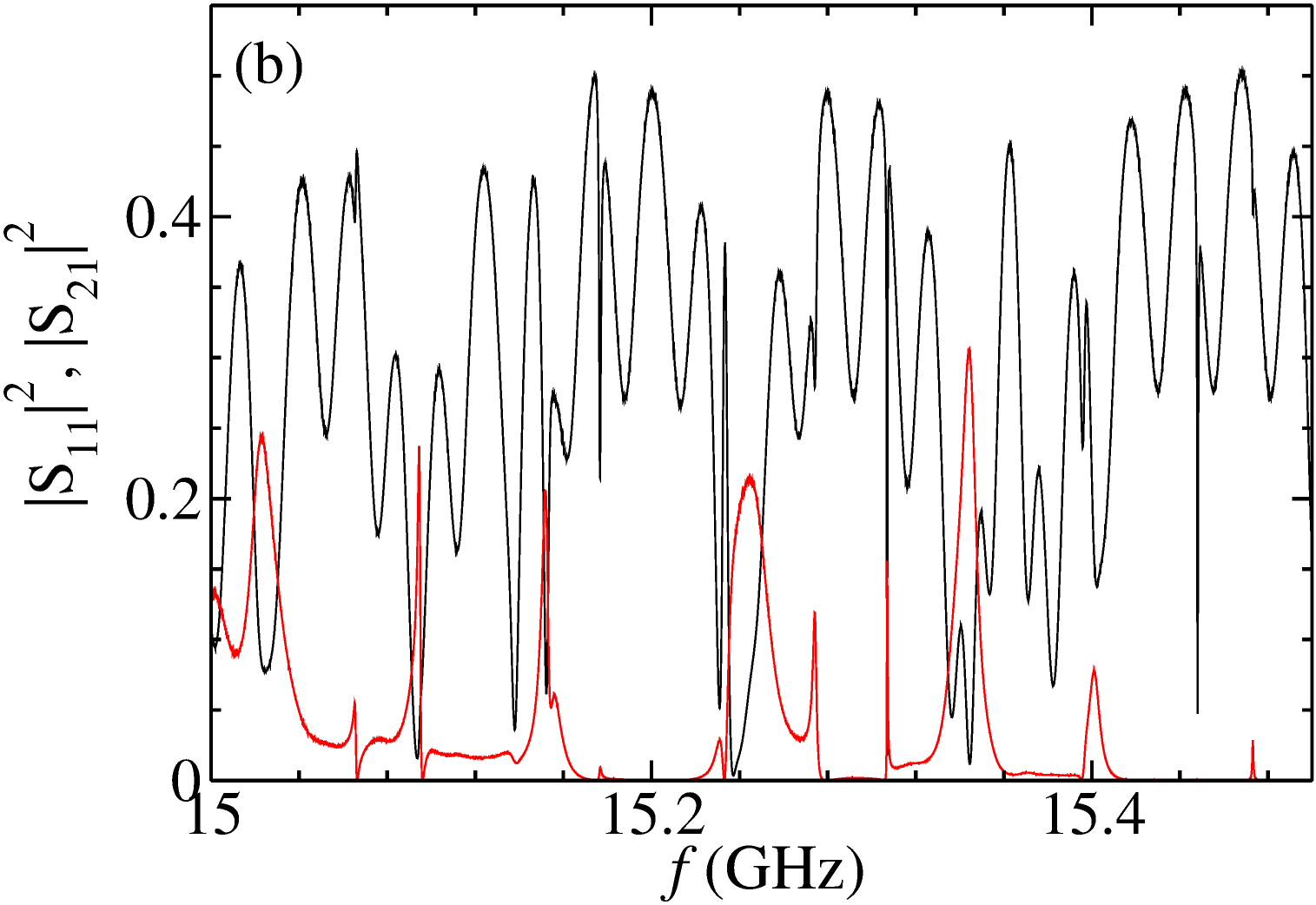}  
\includegraphics[width=0.49\linewidth]{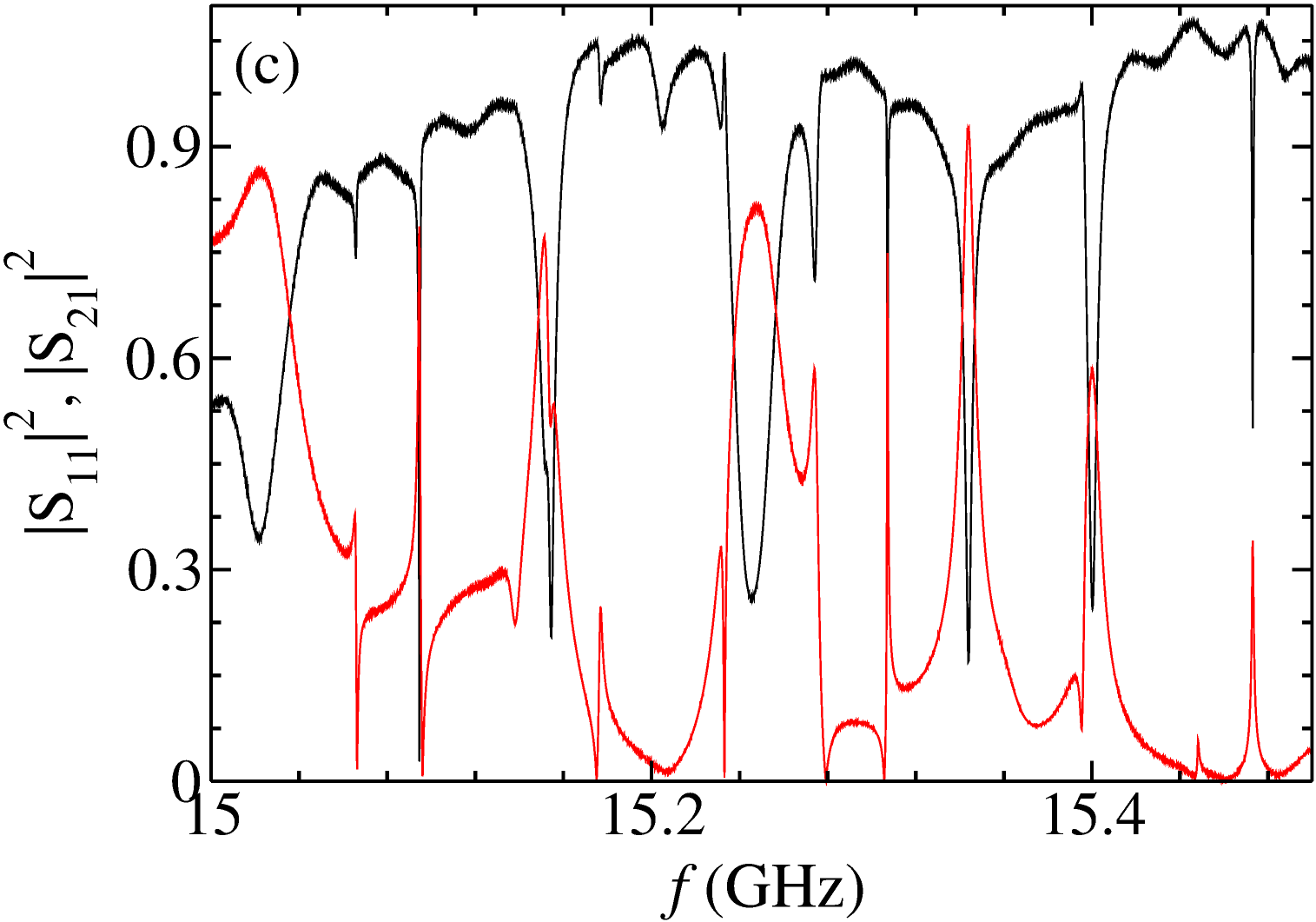}  
\includegraphics[width=0.49\linewidth]{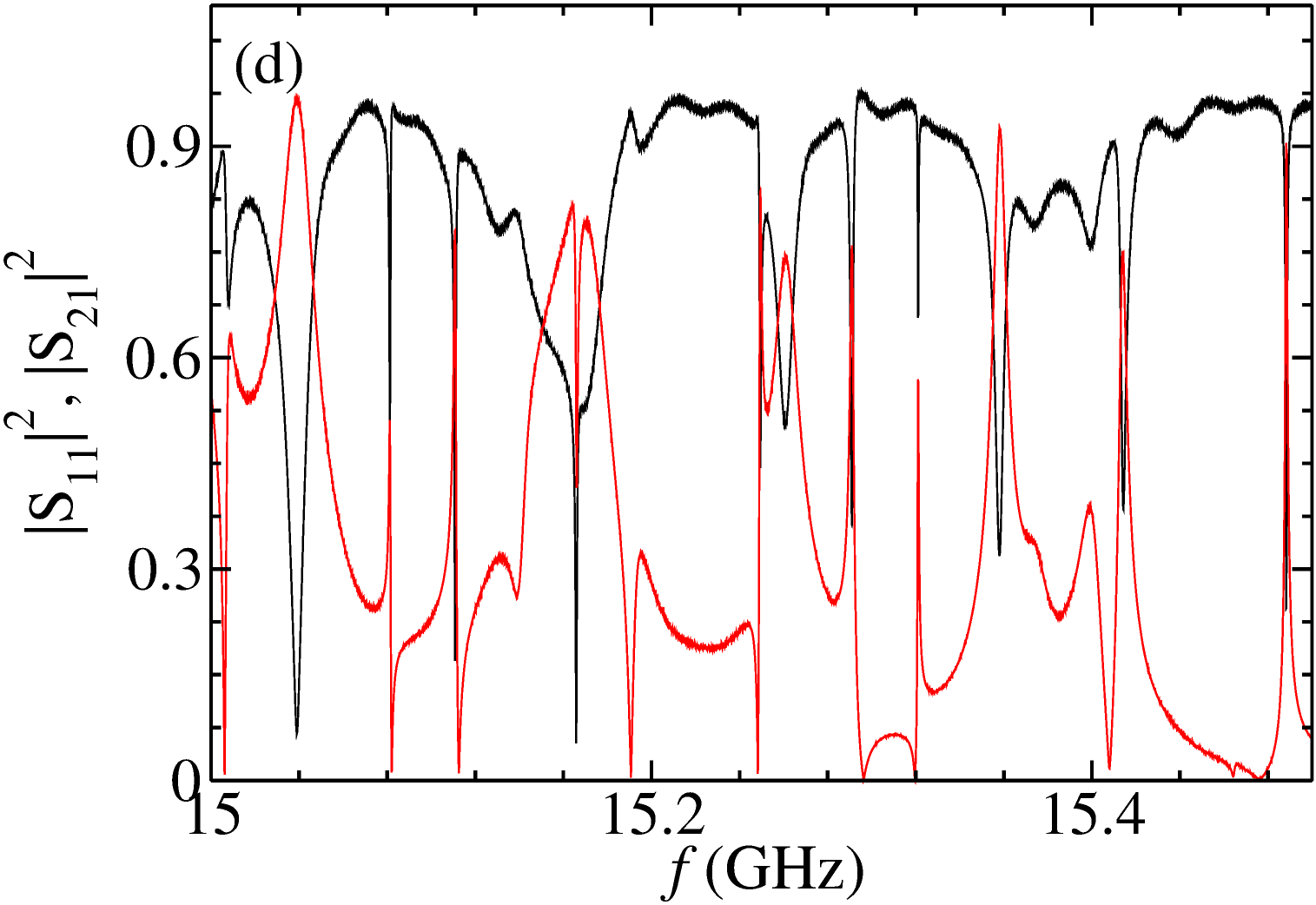}  
	\caption{Top: Uncalibrated reflection (black) and transmission (red) spectra of the tetrahedral (a) and honeycomb (b) waveguide graph. Bottom: Same as top for the calibrated spectra of the tetrahedral (c) and honeycomb (d) waveguide graph.}   
\label{fig:spectra}   
\end{figure}

Figure~\ref{fig:spectra} shows a part of uncalibrated (top) and calibrated (bottom) transmission (red) and reflection (black) spectra measured between two of the three ports for the tetrahedral [(a) and (c)] and honeycomb [(b) and (d)] waveguide graph, respectively. The effect of the TRL calibration is especially visible in the reflection spectra where it removes the oscillations originating from reflections within the cables coupling the VNA to the resonator, clearly visible in the uncalibrated spectra. In the ideal case the reflection spectra should be close to unity away from resonances~\cite{Alt1995}. Yet, we observe in the TRL-calibrated reflection spectra $\vert S_{11}\vert$ a slow oscillation that leads to deviations of $\vert S_{11}\vert$ from unity between resonances and from zero for the corresponding transmission spectra $\vert S_{21}\vert$. These originate from systematic errors in the TRL calibration, that were removed in Ref.~\cite{Yeh2013} by multiplying the $S$ matrix with a frequency-dependent diagonal matrix, which was determined on the basis of an ensemble of similar measurements with the same calibration tool. Such ensembles are not available in our case, but we fit trigonometric functions to the slow oscillations in the reflection spectra to roughly remove them and barely observed changes in the fluctuation properties of the $S$ matrix considered in~\refsec{Smatrix}. 

\section{Results for the measurements with antennas\label{Spectral}}  
For the determination of a complete sequence of eigenfrequencies, a resonator with high quality factor is indispensable, but this is not sufficient, because resonances can not be excited in situations where the electric field strength is zero at the position of an antenna implicating a missing eigenfrequency. To avoid this, we performed measurements for all possible combinations of 8 antennas, of which the positions were distributed over the whole waveguide graph. 

\begin{figure}[htbp]
\includegraphics[width=0.6\linewidth]{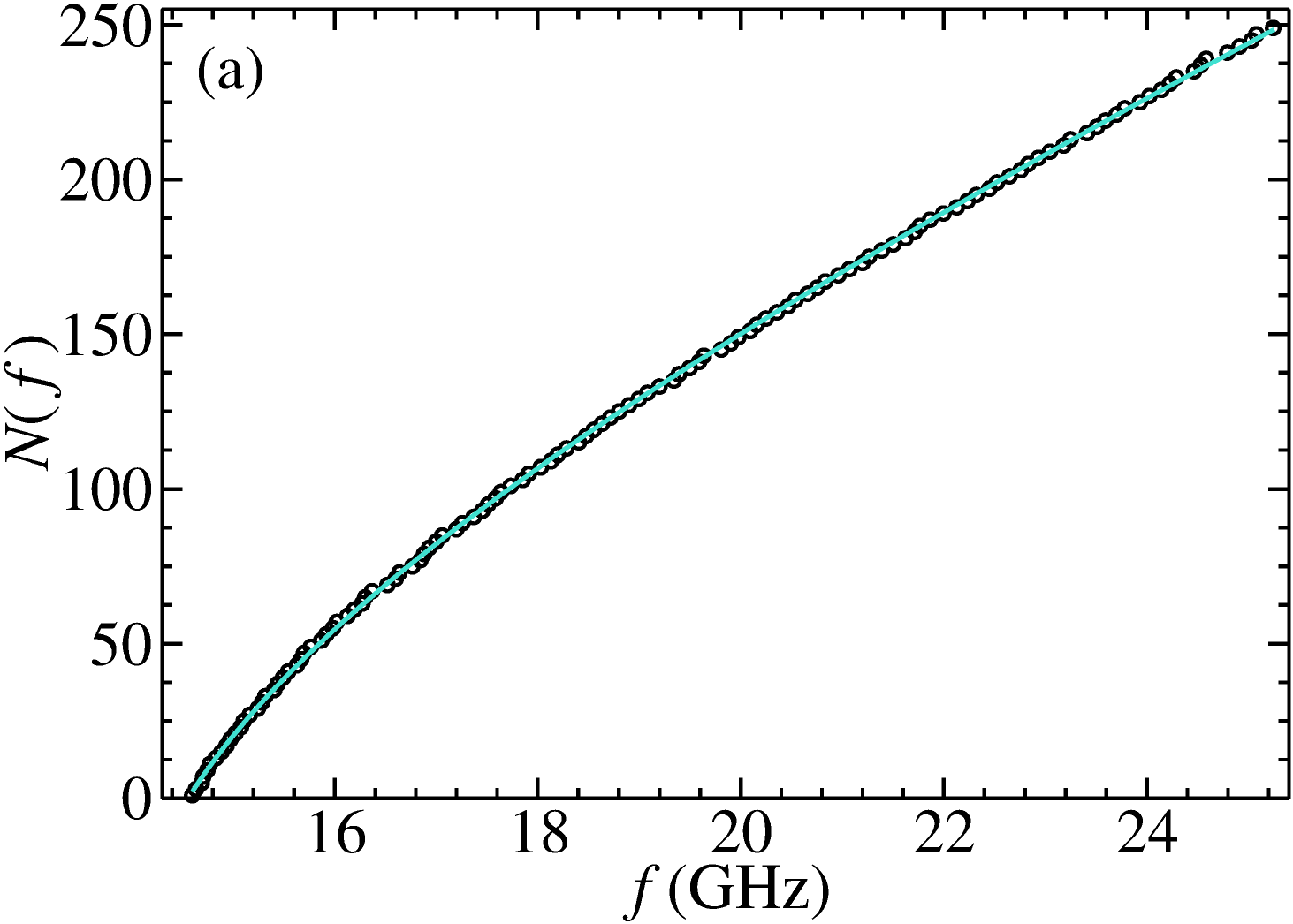}
\includegraphics[width=0.6\linewidth]{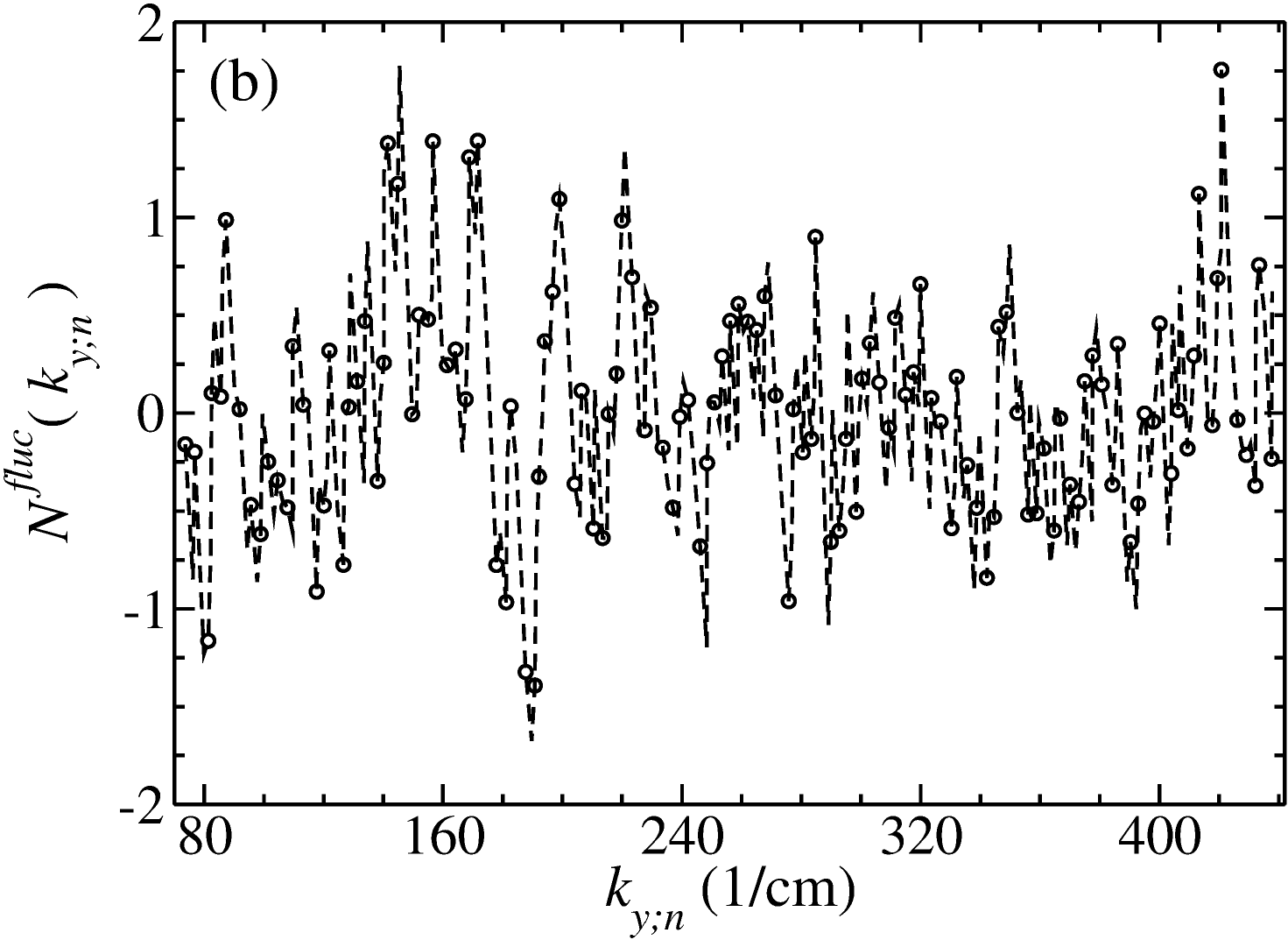}
	\caption{(a) Integrated spectral density deduced from the experimental data (black circles) of the honeycomb waveguide graph in comparison to the analytical results (red solid line), i.e., the smooth part of  $N(f)$ given in~\refeq{Eq:n1}. (b) The fluctuating part of $N(f)$ which is obtained by subtracting the smooth part from it.}
\label{fig:Nf}
\end{figure}

Before analyzing statistical measures for the spectral properties, the eigenfrequencies need to be unfolded to uniform spectral density and mean spacing unity in order to remove system specific properties~\cite{Haake2018}. In distinction to quantum graphs and microwave cable networks the spectral density, and thus the mean spacing depends on the eigenfrequency $f_n$. In the frequency range of a single transversal mode $f_{\rm TM_{10}}\leq f\leq f_{\rm TM_{20}}$ the smooth part of the integrated spectral density is obtained from~\refeq{kym} with $q=1$ and $p=0$,
\begin{equation}\label{Eq:n1}
	N^{smooth}(f_n)=\frac{\mathcal{L}}{\pi}\sqrt{\left(\frac{2\pi f_n}{c}\right)^2-\left(\frac{\pi}{w}\right)^2},
\end{equation}
yielding with~\refeq{kym} $N^{smooth}(k_{y;n})=\frac{\mathcal{L}}{\pi}k_{y;n}$, which is similar to~\refeq{Weyl}. Note, that for frequencies $f_n\gg f_{\rm TM_{10}}$ $N^{smooth}(f)$ approaches that for the corresponding quantum graph or microwave cable network, $N^{smooth}(f_n)\simeq\frac{\mathcal{L}}{\pi}\frac{2\pi f_n}{c}$. Complete sequences of $\approx 250$ eigenfrequencies were identified for both waveguide graphs. In~\reffig{fig:Nf} we compare for the honeycomb waveguide graph the analytical expressions Eqs.~(\ref{Eq:n1}) (turquoise line) to the experimentally obtained integrated spectral density (black circles). Agreement is very good. 
The eigenfrequencies were unfolded to mean spacing unity by replacing them by Eq.~(\ref{Eq:n1}), $\epsilon_n=N^{smooth}(f_n)$. In~\reffig{fig:Nf} we show the fluctuating part of the integrated spectral density, $N^{fluc}(k_{y;n})=N(k_{y;n})-N^{smooth}(k_{y;n})$, which fluctuates around zero, thus indicating that the eigenvalue sequence is complete. Furthermore, we computed 3000 eigenvalues for the corresponding quantum graphs employing the quantization condition~\refeq{QuantCond}. 

\begin{figure}[th!]
\includegraphics[width=0.8\linewidth]{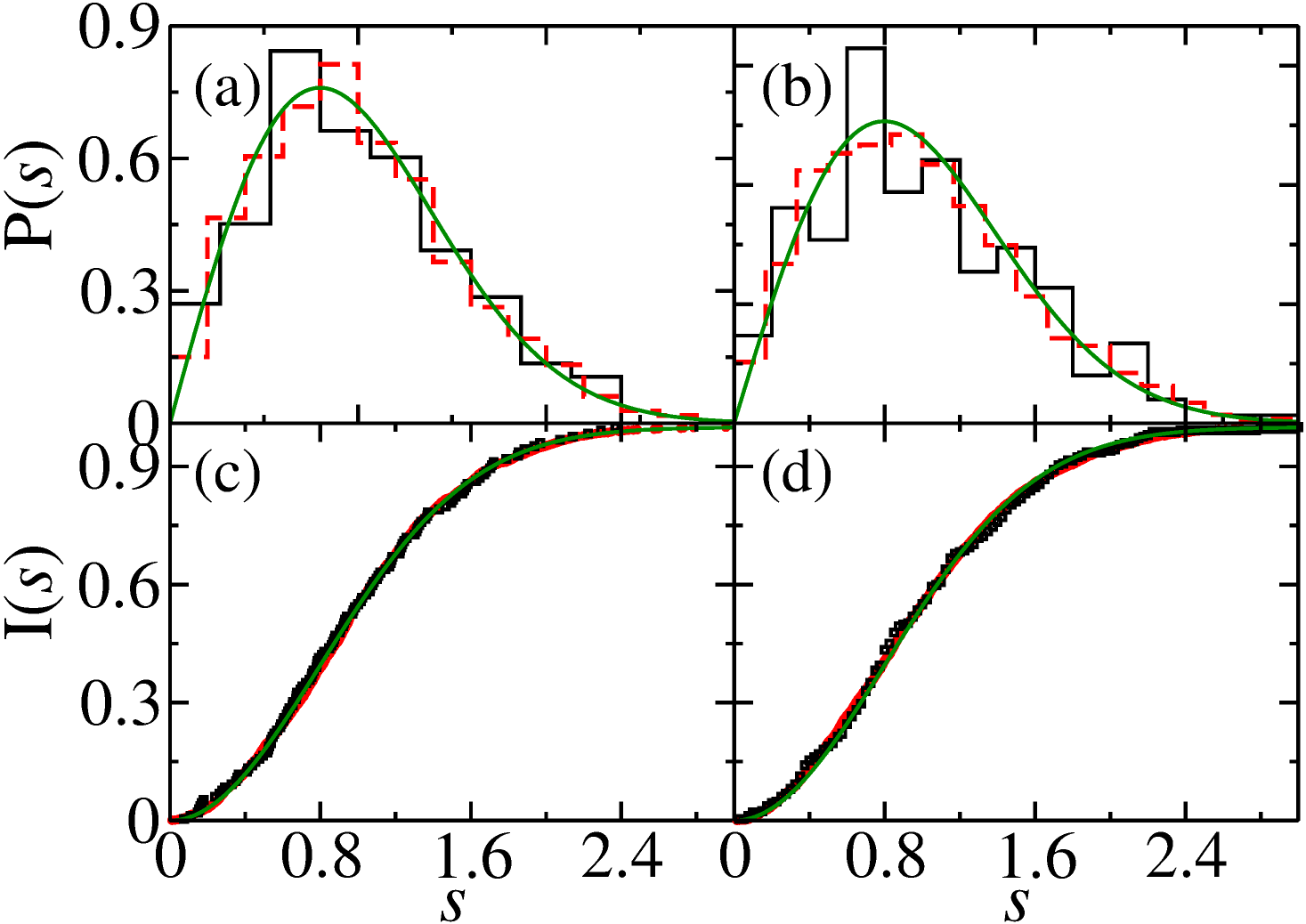}
	\caption{Nearest-neighbor spacing distribution $P(s)$ and cumulative nearest-neighbor spacing distribution $I(s)$ deduced from the unfolded eigenfrequencies of the tetrahedral [(a), (c)] and honeycomb [(b), (d)] waveguide graph (black) and the quantum graph of corresponding geometry (red), respectively. The green full lines show the results for the GOE.}
\label{fig:short}
\end{figure}

\begin{figure}[!th]
\includegraphics[width=0.75\linewidth]{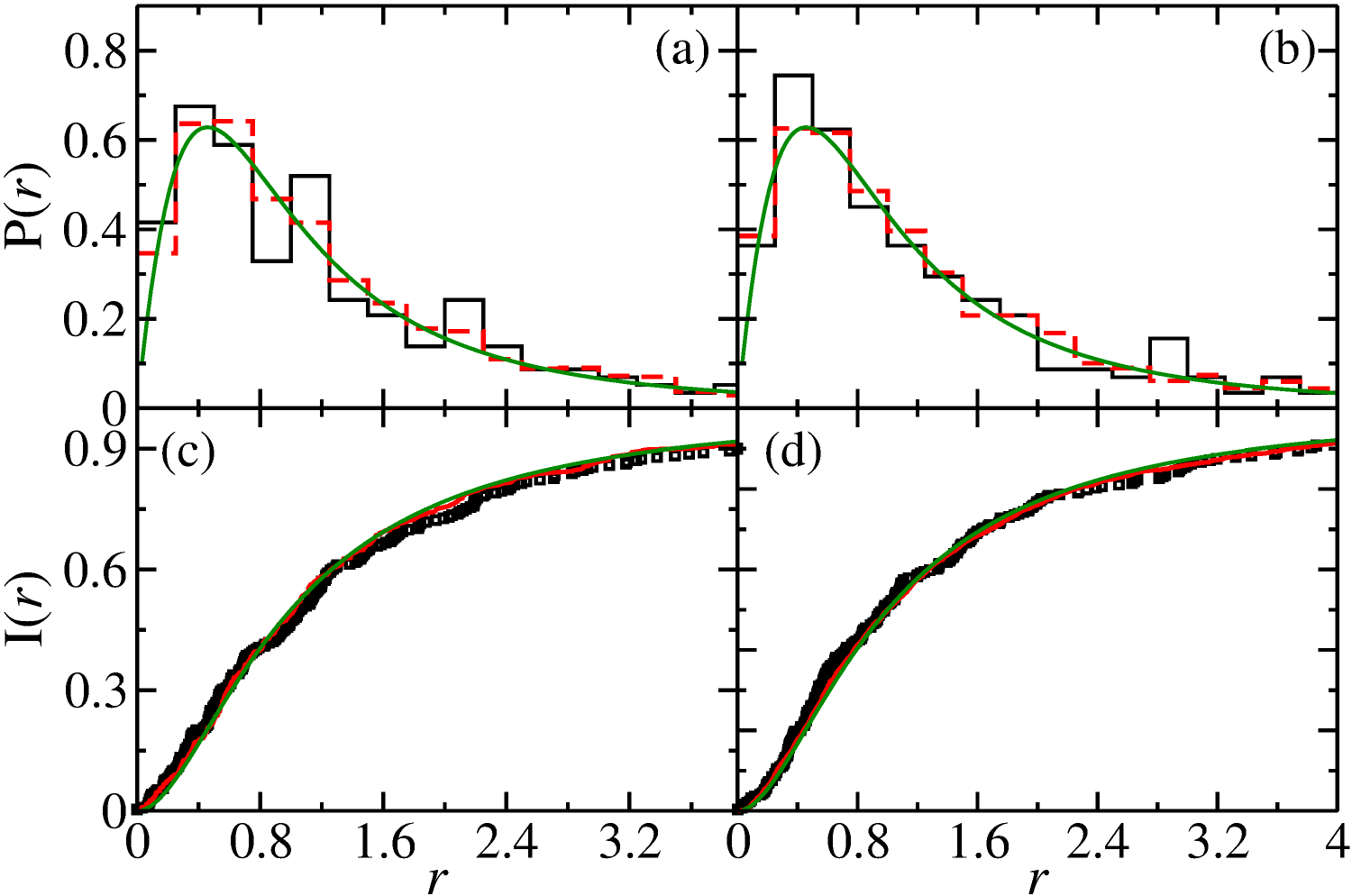}
	\caption{Ratio distributions $P(r)$ and cumulative ratio distributions $I(r)$ deduced from the non-unfolded eigenfrequencies of the tetrahedral [(a), (c)] and honeycomb [(b), (d)] waveguide graph (black) and the quantum graph of corresponding geometry (red), respectively. The green full lines show the results for the GOE.}
\label{fig:ratio}
\end{figure}

\begin{figure}[th!]
\includegraphics[width=0.8\linewidth]{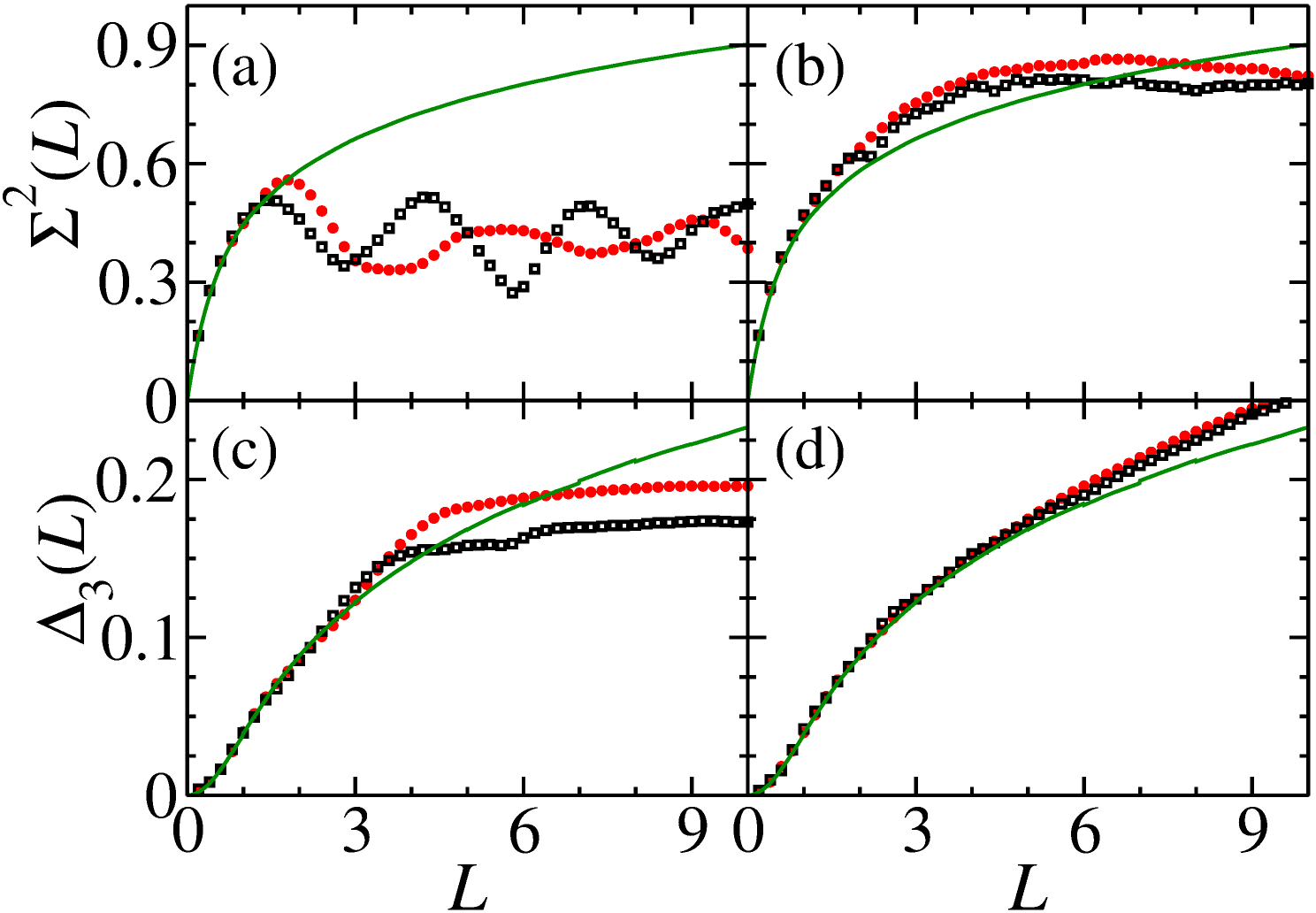}
	\caption{Number variance $\Sigma^2(L)$ and spectral rigidity $\Delta_3(L)$ deduced from the unfolded eigenfrequencies of the tetrahedral [(a), (c)] and honeycomb [(b), (d)] waveguide graph (black) and the quantum graph of corresponding geometry (red), respectively. The green full lines show the results for the GOE.}
\label{fig:long}
\end{figure}

We analyzed spectral properties of the tetrahedral and honeycomb waveguide graphs and compared them to those of the corresponding quantum graphs for a similar number of eigenvalues. The results are shown in the left and right parts of Figs.~\ref{fig:short},~\ref{fig:ratio} and~\ref{fig:long}, respectively. Shown are the distribution of the spacings between nearest-neighbor eigenvalues $P(s)$, the associated cumulative distribution $I(s)$, and the distributions of the ratios of consecutive spacings between next-nearest neighbors $P(r)$~\cite{Oganesyan2007,Atas2013} and the corresponding cumulative distribution $I(r)$ as measures for short-range correlations and the number variance $\Sigma^2(L)$ and Dyson-Mehta statistics $\Delta_3(L)$, that is, the spectral rigidity of the spectra~\cite{Bohigas1975,Mehta1990}, which provide information on long-range correlations. Furthermore, we analyzed the power spectrum which is given in terms of the Fourier transform from $q$ to $\tau$ of the deviation of the $q$th nearest-neighbor spacing from its mean value $q$, $\delta_q=\epsilon_{q+1}-\epsilon_1-q$, 
\begin{equation}
\label{PowerS}
s(\tau)=\left\langle\left\vert\frac{1}{\sqrt{N}}\sum_{q=0}^{N-1} \delta_q\exp\left(-2\pi i\tau q\right)\right\vert^2\right\rangle
\end{equation}
for a sequence of $N$ levels, where $0\leq\tau\leq 1$~\cite{Relano2002,Faleiro2004,Gomez2005,Salasnich2005,Santhanam2005,Relano2008,Faleiro2006,Mur2015}. It was studied experimentally for microwave cable networks with preserved and violated time-reversal invariance and with symplectic symmetry~\cite{Bialous2016,Che2021}.

\begin{figure}[!th]
\includegraphics[width=0.8\linewidth]{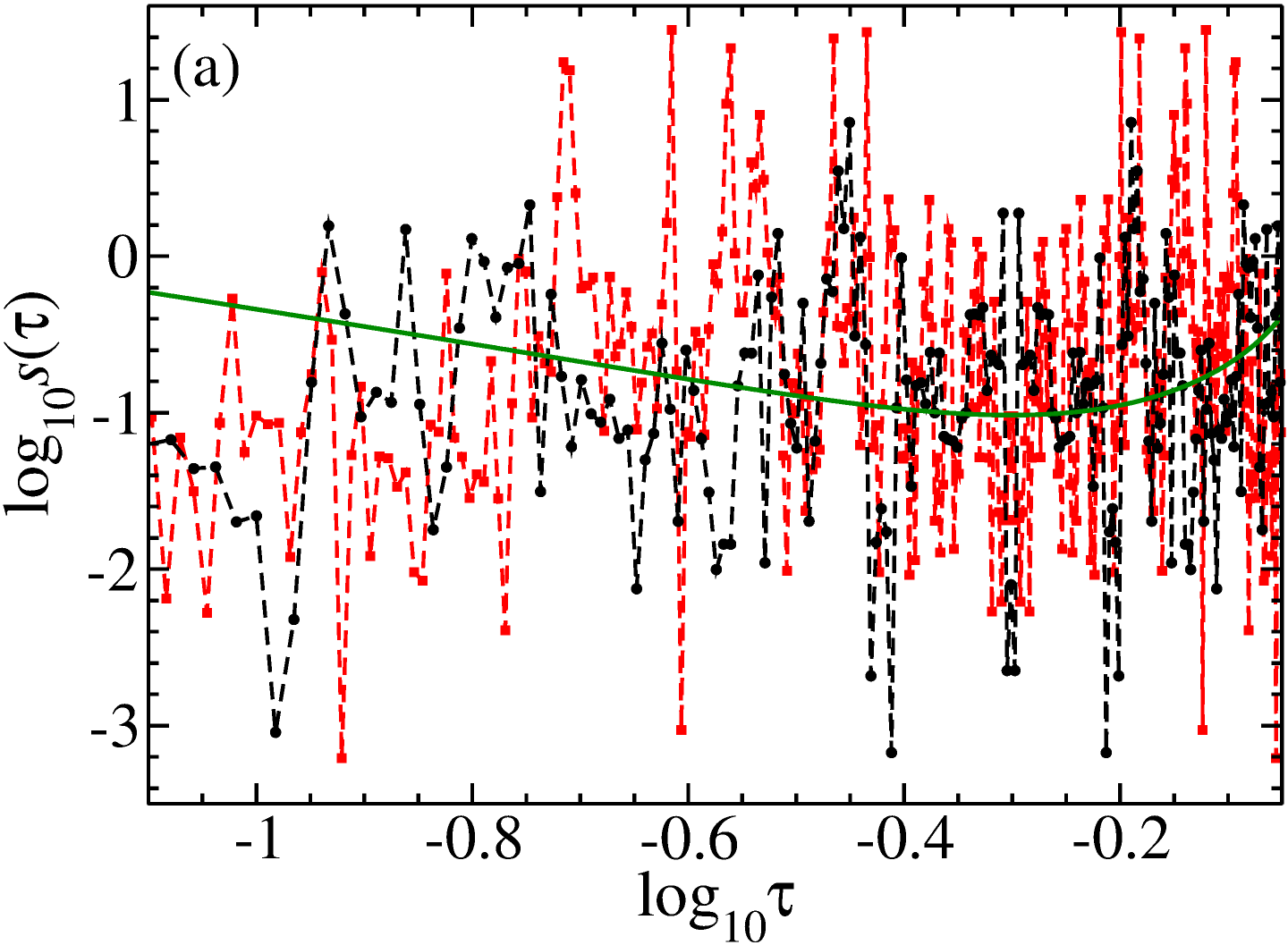}
\includegraphics[width=0.8\linewidth]{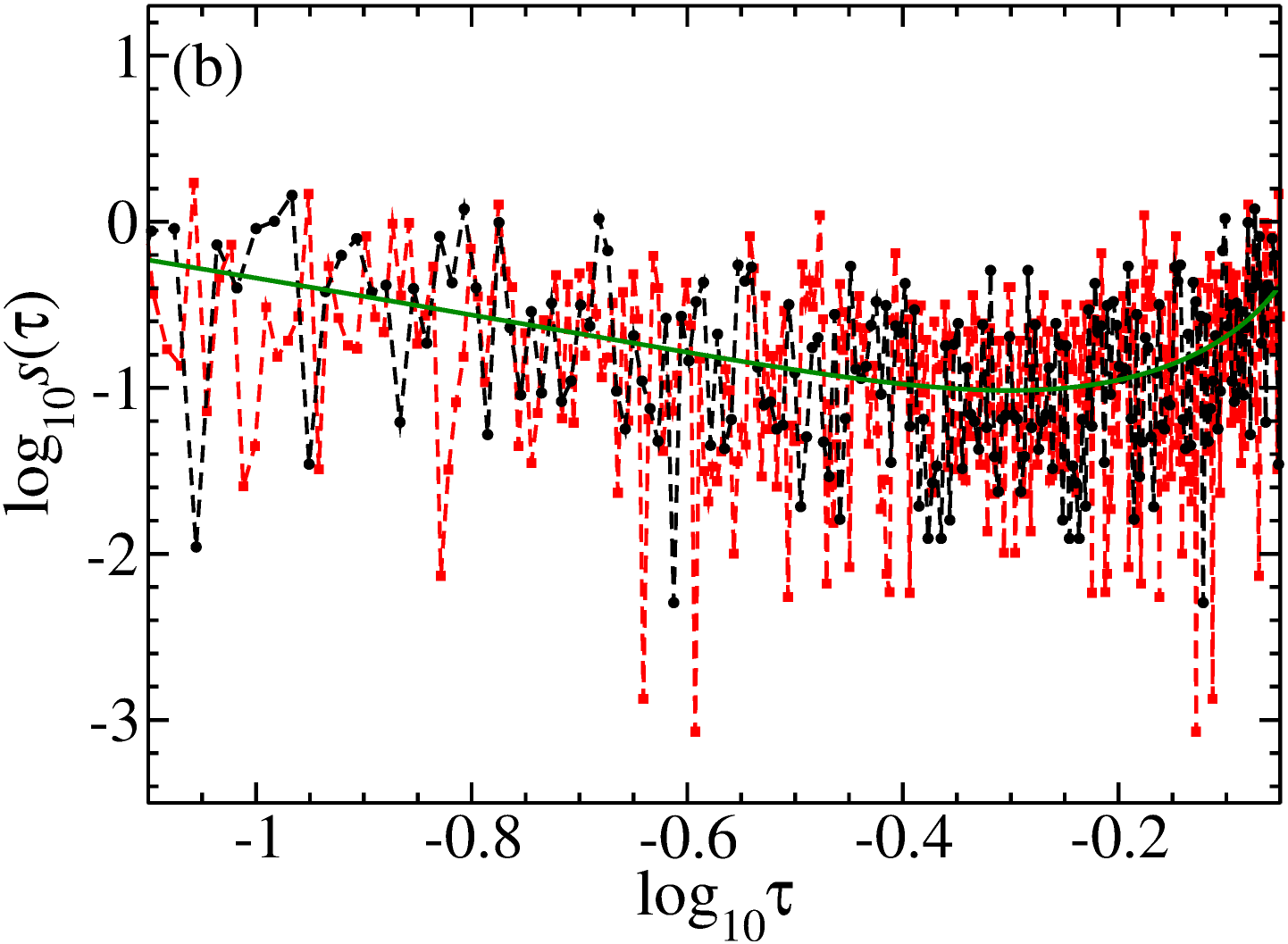}
\caption{Comparison of the power spectra of the tetrahedral (a) and honeycomb (b) waveguide graphs (black) and corresponding quantum graphs (red). The green lines exhibit the results for random matrices from the GOE.}
\label{fig:power}
\end{figure}

Agreement with the curves for GOE statistics are good for both waveguide graphs and the corresponding quantum graphs as concerns short range correlations; see Figs.~\ref{fig:short} and~\ref{fig:ratio}. However, deviations are clearly visible in Figs.~\ref{fig:long} and~\ref{fig:power} for the long range correlations for the tetrahedral waveguide and quantum graphs shown in (a) and (c), especially for $\Sigma^2(L)$  beyond $L\gtrsim 2-3$. These deviations may be attributed to the small number of vertices, an inappropriate choice of the relative lengths, which should be incommensurable, and/or backscattering at the vertices~\cite{Dietz2017b}, which leads to their confinement to individual bonds or a fraction of them~\cite{Dietz2017b}. The oscillations observed in $\Sigma^2(L)$ for the waveguide graph are due to a bad choice of bond lengths, but since they are not that pronounced for the quantum graph, may also be attributed to the stronger backscattering at the $90^\circ$ bends and T junctions~\cite{Bittner2013} illustrated in Figs.~\ref{SmatrWG} and~\ref{SmatrYT}. Similar discrepancies are observed for the power spectrum in~\reffig{fig:power} where large fluctuations about the RMT prediction are observed for the tetrahedral waveguide and quantum graphs (a) for ${\rm log}_{10}\tau\gtrsim -1.0$ and deviations below that value. These are slightly larger for the waveguide graph. For the honeycomb waveguide graph and corresponding quantum graph, shown in (b), on the contrary, they are similar and agree well down to ${\rm log}_{10}\tau\approx -1.3$, even though the number of the eigenfrequency sequences is for both types of graphs the same as for the tetrahedral geometry. Note, that the design of the first waveguide graph, which was the tetrahedral graph, was not optimized, whereas we performed numerous numerical simulations with CST Studios with varying number of vertices and also with T joints, before we constucted the honeycomb waveguide graph. The agreement of the spectral properties of the honeycomb waveguide graph with RMT predictions is good and slightly better than for the corresponding quantum graph, thus indicating that it provides a suitable experimental system for the study of quantum-chaotic or wave-chaotic features~\cite{Zhang2022}. In the waveguide graphs, microwave cable networks and quantum graphs the wave propagation is one dimensional along the bonds, but the complexity is induced by the transport properties at the junctions, corroborating their crucial role in the design of wave-dynamical features~\cite{Gnutzmann2006}.  

\begin{figure}[htbp]
\includegraphics[width=0.9\linewidth]{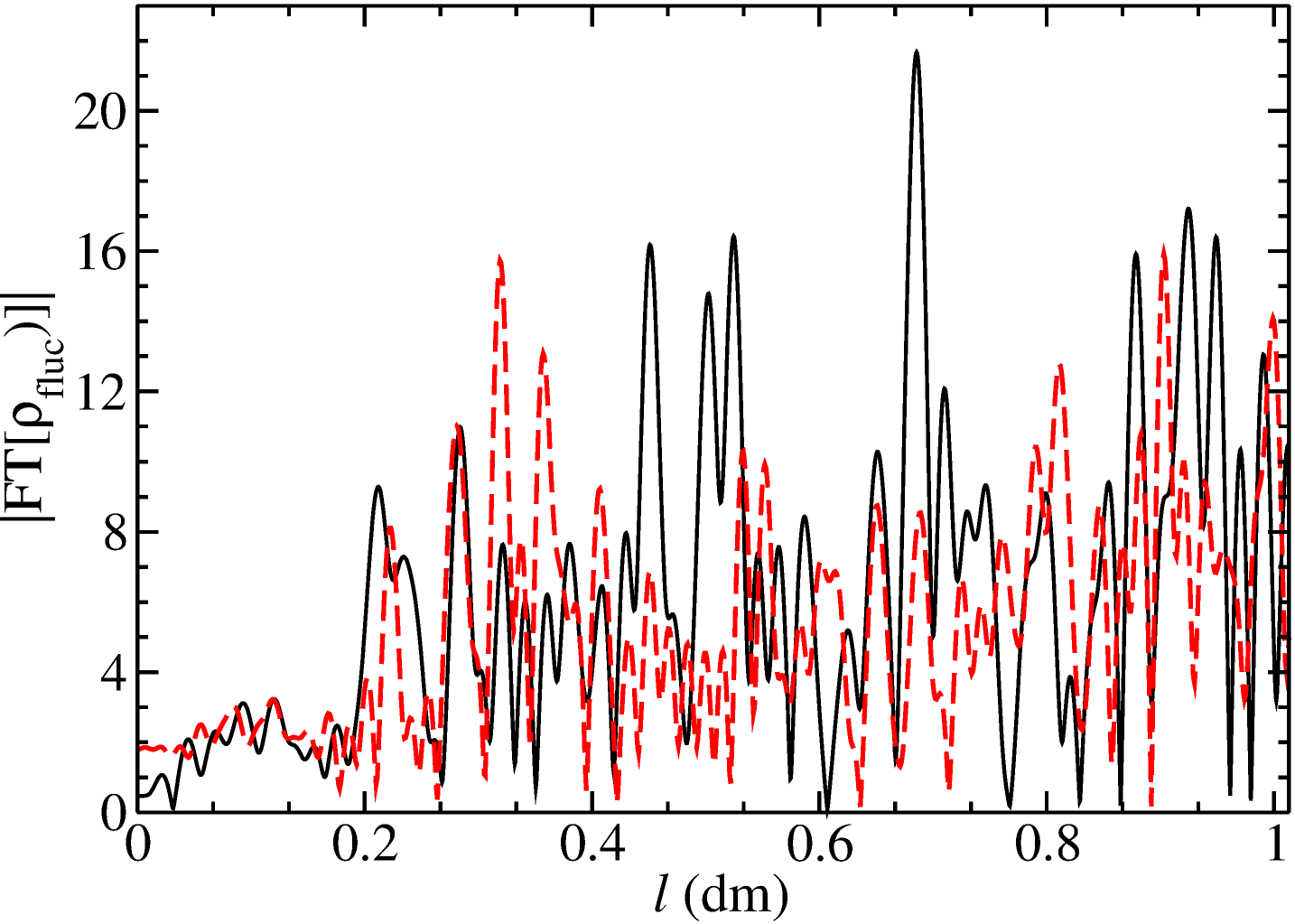}  
\caption{Comparison of the length spectrum of the honeycomb waveguide graph (black dashed line) with that obtained from the lowest 250 eigenvalues of the quantum graph of corresponding geometry (red line).}  
\label{fig:ls}  
\end{figure}

In~\reffig{fig:ls} we compare length spectra deduced from the eigenfrequencies of the honeycomb waveguide graph (black solid line) with those obtained from the eigenvalues of the quantum graph of corresponding geometry, where we took into account a similar number of eigenvalues (red dashed lines). A length spectrum corresponds to the modulus of the Fourier transform of the fluctuating part of the spectral density from wavenumber to length and has the property that it exhibits peaks at the lengths of the periodic orbits of the corresponding classical system. In a quantum graph periodic orbits consist of trajectories along successive bonds that are uniquely defined by the sequence of the vertices connecting them. Similarly, in the waveguide graph the orbits correspond to the paths of the waves through the waveguide network. The length spectra exhibit peaks at similar lengths, but differ in amplitude, which may be attributed to the distinct features of the vertex $S$ matrices defining the wave propagation through the junctions. We defined the lengths of the waveguide graphs as the distance between the centers of the junctions terminating them, so that they may differ from those of the quantum graph. Furthermore, bends connecting two waveguides correspond to vertices in the waveguide graph and lead to additional interference effects and peaks in their length spectra.  

\section{Scattering experiments with the open honeycomb waveguide graph\label{Smatrix}}

\begin{figure}[!ht]
\includegraphics[width=0.9\linewidth]{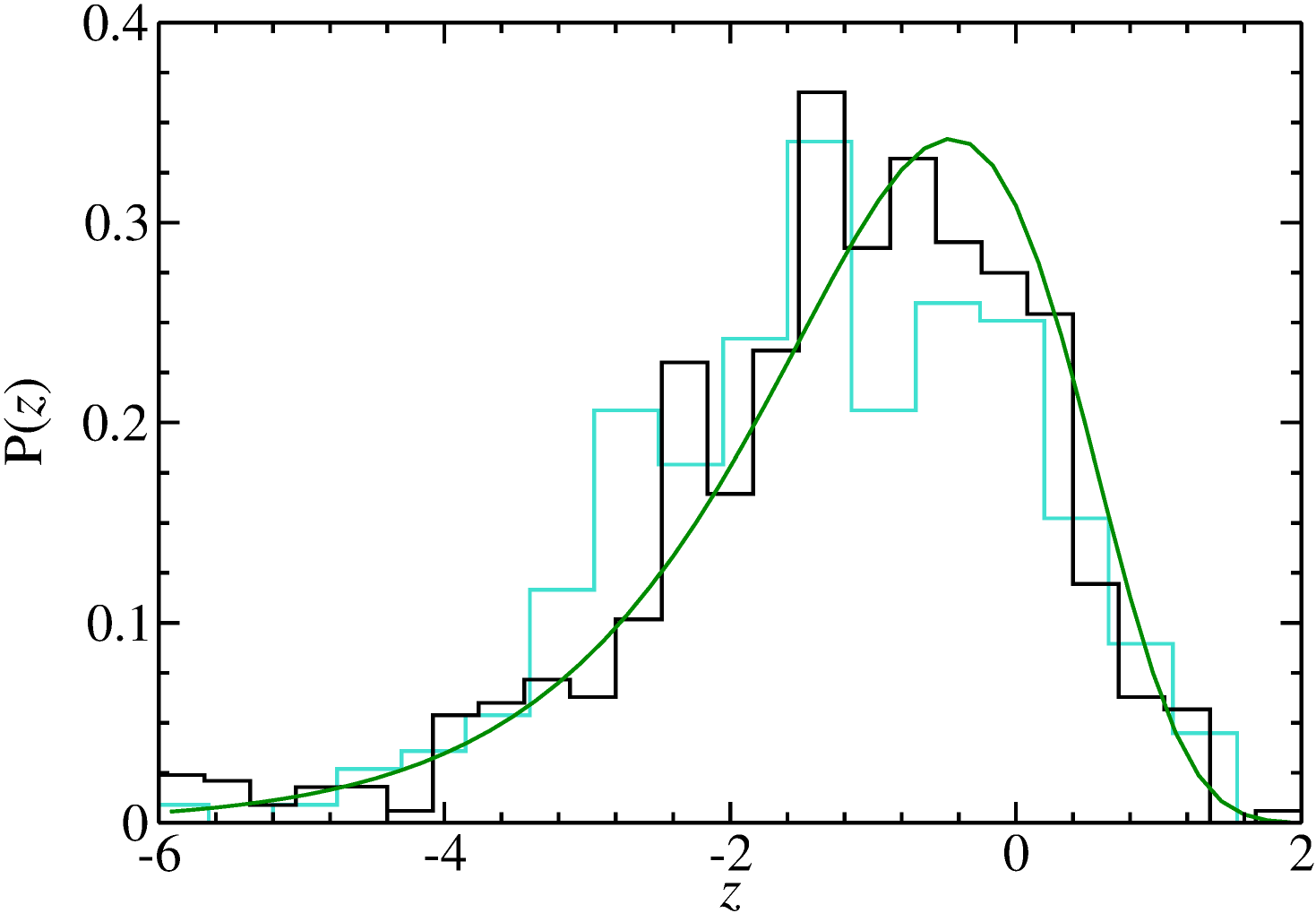}
\caption{Strength distribution. The black histogram shows the result deduced from the the calibrated spectra of the honeycomb waveguide graph, the turquoise histogram exhibits that deduced from the uncalibrated spectra and the green solid line the RMT prediction given in~\refeq{K0}.}
\label{fig:K0}    
\end{figure}

We investigated the distribution of the resonance strengths $y=\gamma_{\mu a}\gamma_{\mu b}$. These were obtained from the fit of~\refeq{Eq:bw} to the resonances in the calibrated transmission spectra. We rescaled them to average value unity, $\langle\tilde y\rangle=1$ and analyzed their distribution. It is compared in~\reffig{fig:K0} for the transformed strength, $z=\log_{10}\widetilde{y}$, with that expected for typical quantum systems with a fully chaotic classical counterpart~\cite{Dembowski2005},
\begin{equation}
\label{K0}
p(z)=\frac{\ln(10)}{\pi}10^{z/2}K_0(10^{z/2}).
\end{equation}
Here, $K_0(x)$ denotes the modified Bessel function of order zero~\cite{Abramowitz2013}. Agreement with the RMT prediction is good. Note, that deviations are expected for the wave functions of quantum graphs~\cite{Kaplan2001,Zhang2022}, however, in the evaluation of their properties in terms of the strength distributions only their values at the positions of the ports are taken into account. Deviations observed in the distribution of the resonance strengths of the closed honeycomb waveguide graph are actually similar to those observed in~\cite{Dembowski2005} for a superconducting microwave billiard with mixed regular-chaotic dynamics. Note that in Ref.~\cite{Dembowski2005} the resonance strengths were obtained by fitting~\refeq{Eq:bw} to resonances in the uncalibrated spectra, however the quality factor of these microwave billiards was by a factor of $\gtrsim 10$ higher than in the present case. The turquoise curve shows the strength distribution deduced from the uncalibrated spectra of the honeycomb waveguide graph. It deviates from that for the calibrated spectra, thus corroborating that calibration is essential.

The $S$ matrix describing that of a microwave resonator is given in~\refeq{eq:SResonator}. It has the same form as the $S$ matrix of the Heidelberg approach which was introduced in~\cite{Mahaux1969} for the description of compound-nucleus reactions. The analogy has been exploited to study aspects of quantum-chaotic scattering in numerous experiments~\cite{Bluemel1990,Kuhl2005,Dietz2008,Chen2021,Che2022}, also for systems with partially or completely violated time-reversal invariance~\cite{Dietz2009,Dietz2010,Lawniczak2020,Zhang2022}, where analytical results were verified~\cite{Verbaarschot1985,Pluhar1995,Fyodorov2005,Dietz2009,Kumar2013,Kumar2017} that had been derived based on the supersymmetry and RMT approach.

The $S$ matrix of the Heidelberg approach is obtained by replacing in~\refeq{eq:SResonator} the resonator Hamiltonian by a random matrix from the GOE, and the matrix elements of the real matrix $\hat W$ by Gaussian distributed random numbers with zero mean value subject to the orthogonality condition $\sum^N_{\mu=1}W_{c\mu}W_{c^\prime\mu} = Nv^2_c\delta_{cc^\prime}$, which expresses the property that the average $S$ matrix is diagonal in the experiments, $\langle S_{ab}\rangle=\delta_{ab}\langle S_{aa}\rangle=\delta_{ab}\frac{(1-\pi^2v_a^2/d)}{(1+\pi^2v_a^2/d)}$ with $d=\sqrt{\frac{2}{N}\langle H^2_{\mu\mu}\rangle}\frac{\pi}{N}$ denoting the mean resonance spacing~\cite{Dietz2010}, implying that there are no direct processes. The matrix $\hat W$ accounts for the coupling of the $N$ internal modes to their environment through $\tilde M$ open channels $c=1,2,\dots,\tilde M$. In the experiments these are $M=3$ open channels corresponding to the three ports. To account for the absorption into the walls of the resonator either an absorption parameter $\tau_{abs}$ may be introduced~\cite{Savin2006,Ericson2013}, or $\Lambda$ identical fictitious channels~\cite{Verbaarschot1986,Dietz2010} yielding $\tilde M=M+\Lambda$. We compared the experimental results to analytical ones, of which the input parameters are the transmission coefficients associated with the ports,
\begin{equation}\label{eq:transcoe}
T_a=1-\vert\langle S_{aa}\rangle\vert^2,\, a=1,2,3
\end{equation}
that provide a measure for the unitarity deficit of the average scattering matrix $\langle S_{aa}\rangle$. They are determined from the measured reflection spectra. The absorption is quantified in terms of the parameter $\tau_{abs}$ which, when ficitious channels are introduced is related to it through $\tau_{abs}=\Lambda T_f$, with $T_f$ denoting the transmission coefficients of the fictitious channels, and can be estimated from the Weisskopf formula 
\be
2\pi\frac{\Gamma}{d}=T_1+T_2+T_3+\tau_{abs},
\ee
where $\Gamma$ is the average resonance width, $\Gamma =\frac{1}{N}\sum_{n=1}^N\Gamma_n$, with $N$ denoting the number of resonances. We obtained an estimate for the transmission coefficients and absorption coefficient from the experimental data and refined them by comparing experimental results for $S$-matrix fluctuation properties to analytical ones~\cite{Dietz2010}.
 
We analyzed the two-point $S$-matrix correlation function and compared it to the analytical result~\cite{Verbaarschot1985},
\begin{equation}\label{eq:autocor}
C_{ab}(\varepsilon)=\langle S^{\rm fl}_{ab}(f)S^{\ast\rm fl}_{ab}(f+\varepsilon)\rangle
\end{equation}
with $S^{\rm fl}_{ab}(f)=S_{ab}(f)-\langle S_{ab}(f)\rangle$ and $\langle\cdot\rangle$ denoting the spectral average over a measured resonance spectrum and an ensemble average over the different port measurements. The analysis was performed in 1~GHz windows to ensure that the resonance parameters are approximately frequency independent, as assumed in the Heidelberg approach for the coupling matrix $\hat{W}$. As in the experiments with a waveguide graph at room temperature~\cite{Zhang2022}, the transmission coefficients obtained in the port measurements are considerably larger than those for the antenna measurements, which is expected since, as outlined in~\refsec{Exp}, the ports couple the resonator modes stronger to the exterior than the antennas. Furthermore, the transmission coefficients and $\tau_{abs}$ barely depend on frequency so that we could analyze the data in the whole available frequency range, where TRL calibration is applicable~\cite{Yeh2013}. We chose the lengths of the two Line standards in~\reffig{fig:Cryostat} such that this was the case for the range $14.5 -19.5$~GHz and $21-26.5$~GHz. 

\begin{figure}[htbp]
\includegraphics[width=0.7\linewidth]{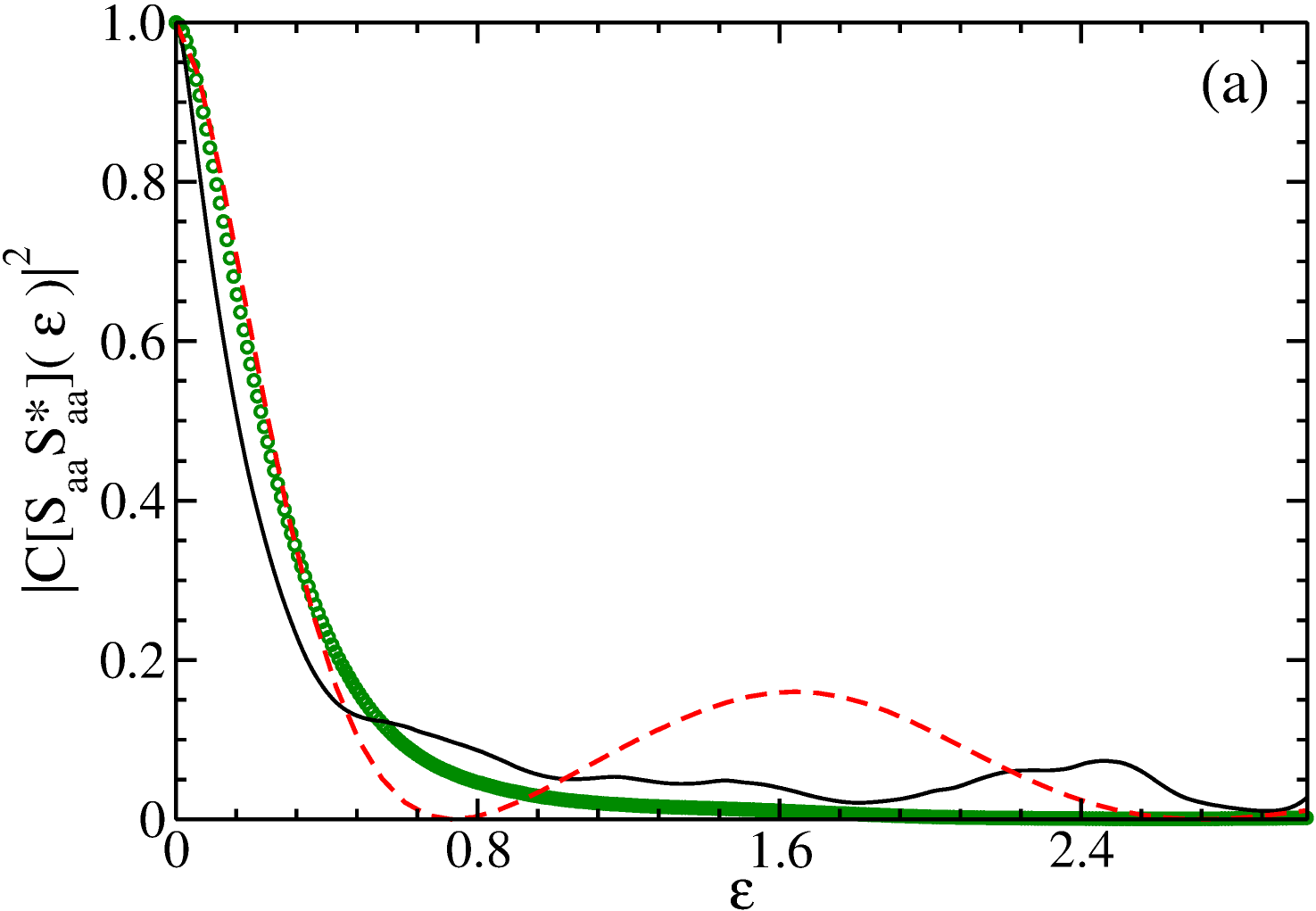}  
\includegraphics[width=0.7\linewidth]{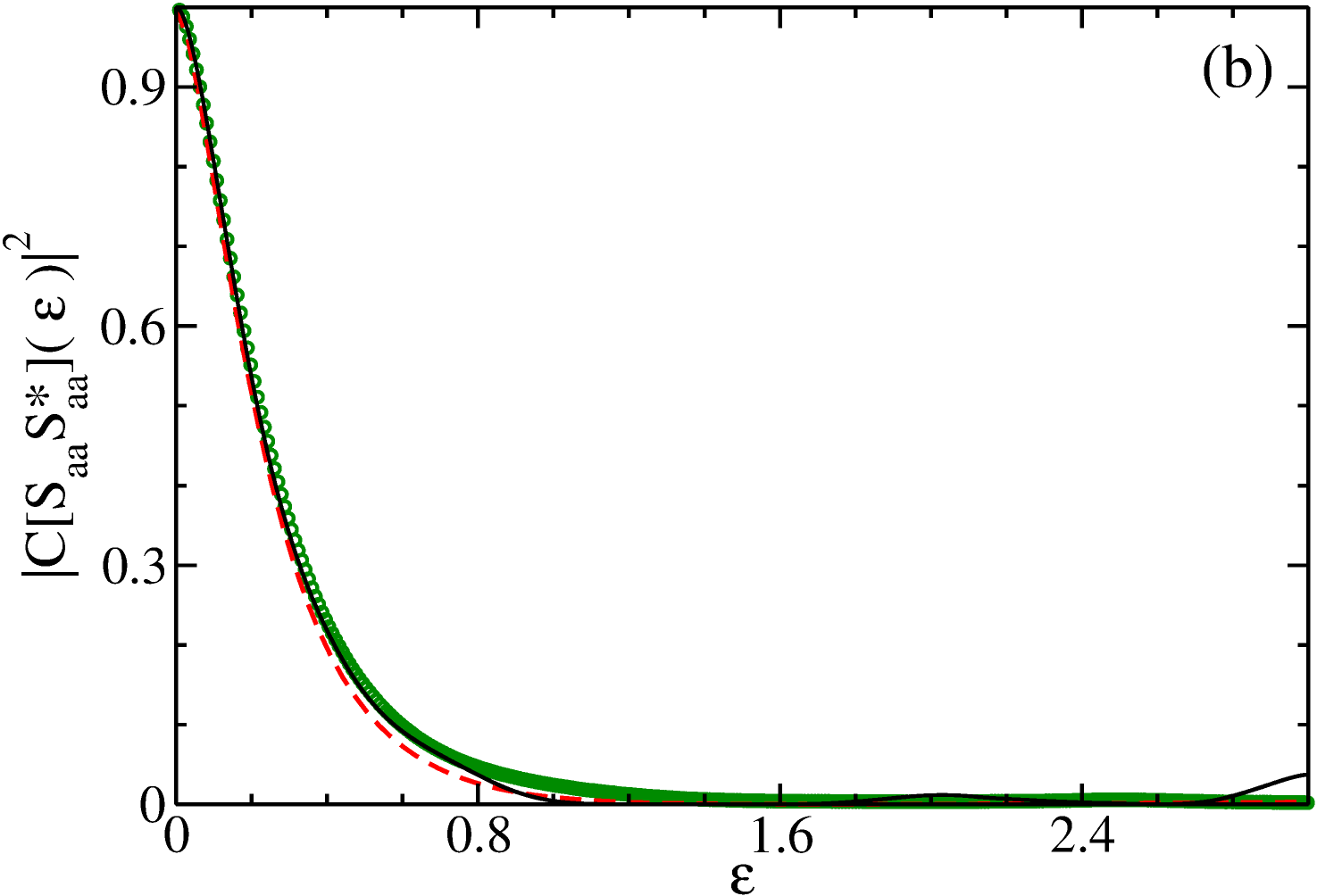}  
	\caption{Two-point $S$-matrix correlation functions obtained from the port measurement with a=b=1 (black lines) and from numerical computations for the corresponding open quantum graph (red) for (a) the tetrahedral graph and (b) the honeycomb graph. They are compared to the RMT results (green circles) for the values of $T_1=T_2=T_3=0.9$ and $T_1=T_2=T_3=0.85$, respectively, with $\tau_{abs}=0.01$. The frequency difference $\epsilon$ is plotted in units of the local mean level spacing $d$. All correlation functions are scaled such that they equal unity at $\epsilon = 0$.}  
\label{fig:corrport}   
\end{figure}

In~\reffig{fig:corrport} (a) we show the two-point $S$-matrix correlation functions of the tetrahedral waveguide (black solid line) and quantum (red dashed line) graphs. Only the latter agrees well with the corresponding RMT curve (green circles), whereas agreement between the experimental curve and the corresponding RMT result is very good for the honeycomb waveguide as demonstrated in~\reffig{fig:corrport} (b). The same applies to the curve for the corresponding open quantum graph (red dashed line) which is nearly on top of that for the waveguide graph, thus corroborating that the effect of absorption into the walls of the waveguide graph indeed is negligible. The values of the transmission coefficients and  the absorption parameter $\tau_{abs}$ are given in the figure caption. The size of $\tau_{abs}$ is indeed small compared to that of the transmission coefficients and to typical values obtained in room temperature measurements in that frequency range~\cite{Dietz2010,Zhang2022}. 

\begin{figure}[htbp]
\includegraphics[width=0.7\linewidth]{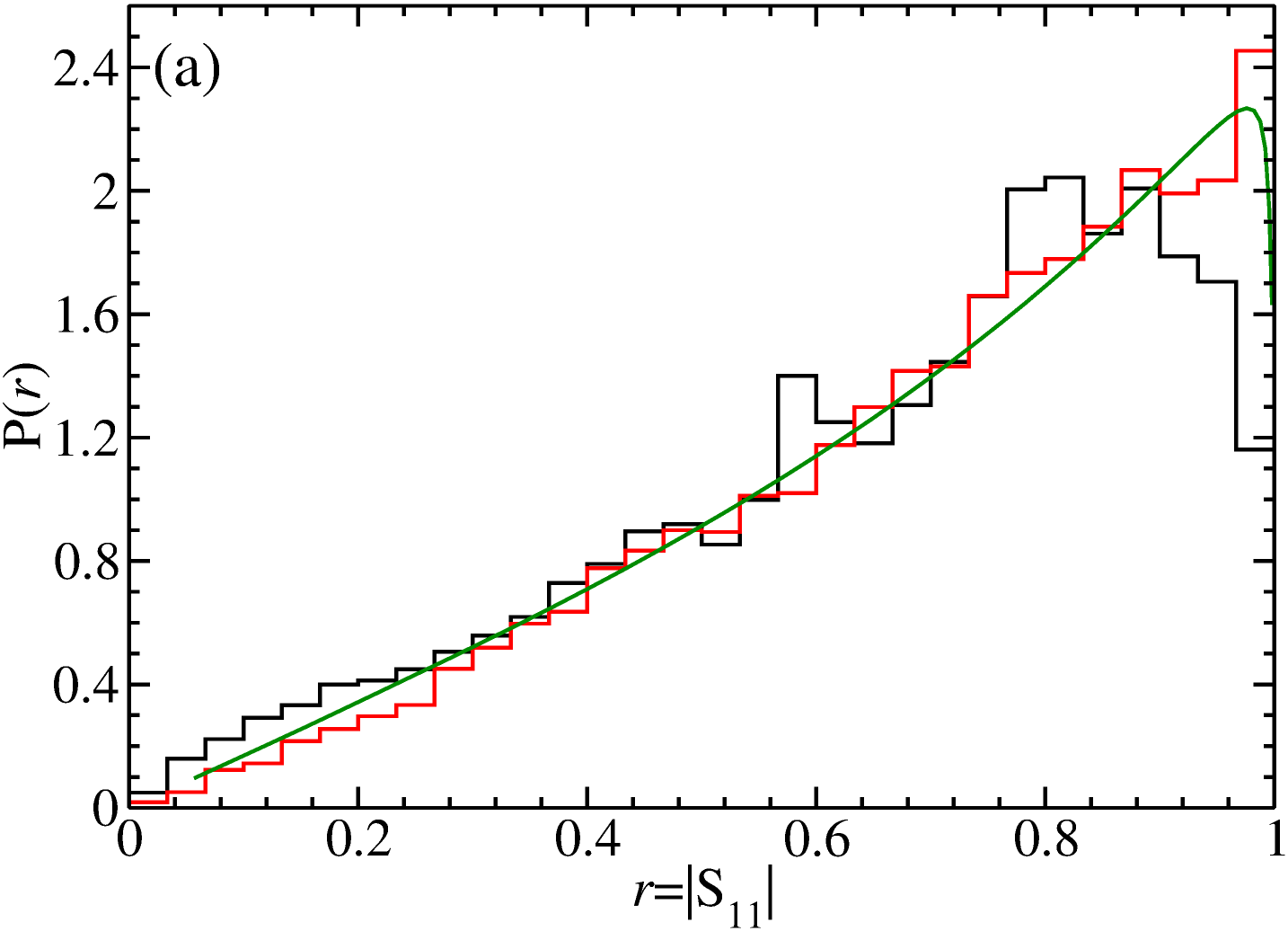}
\includegraphics[width=0.7\linewidth]{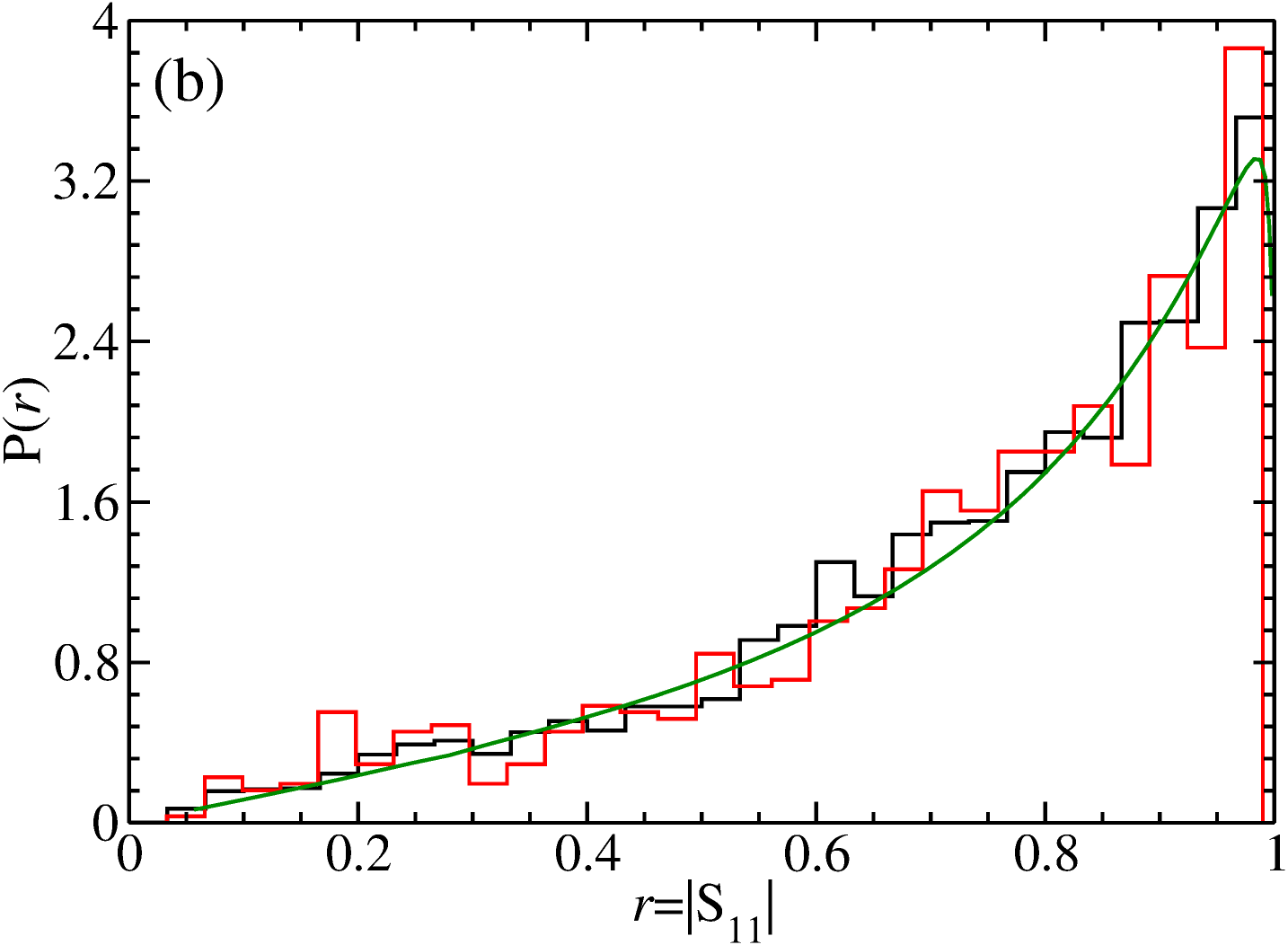}
	\caption{Same as ~\reffig{fig:corrport} for the distribution of the rescaled modulus of $S$-matrix elements, $\tilde r= \vert S_{11}\vert /\langle\vert S_{11}\vert\rangle$.}
\label{fig:s12disport}
\end{figure}

\begin{figure}[htbp]
\includegraphics[width=0.7\linewidth]{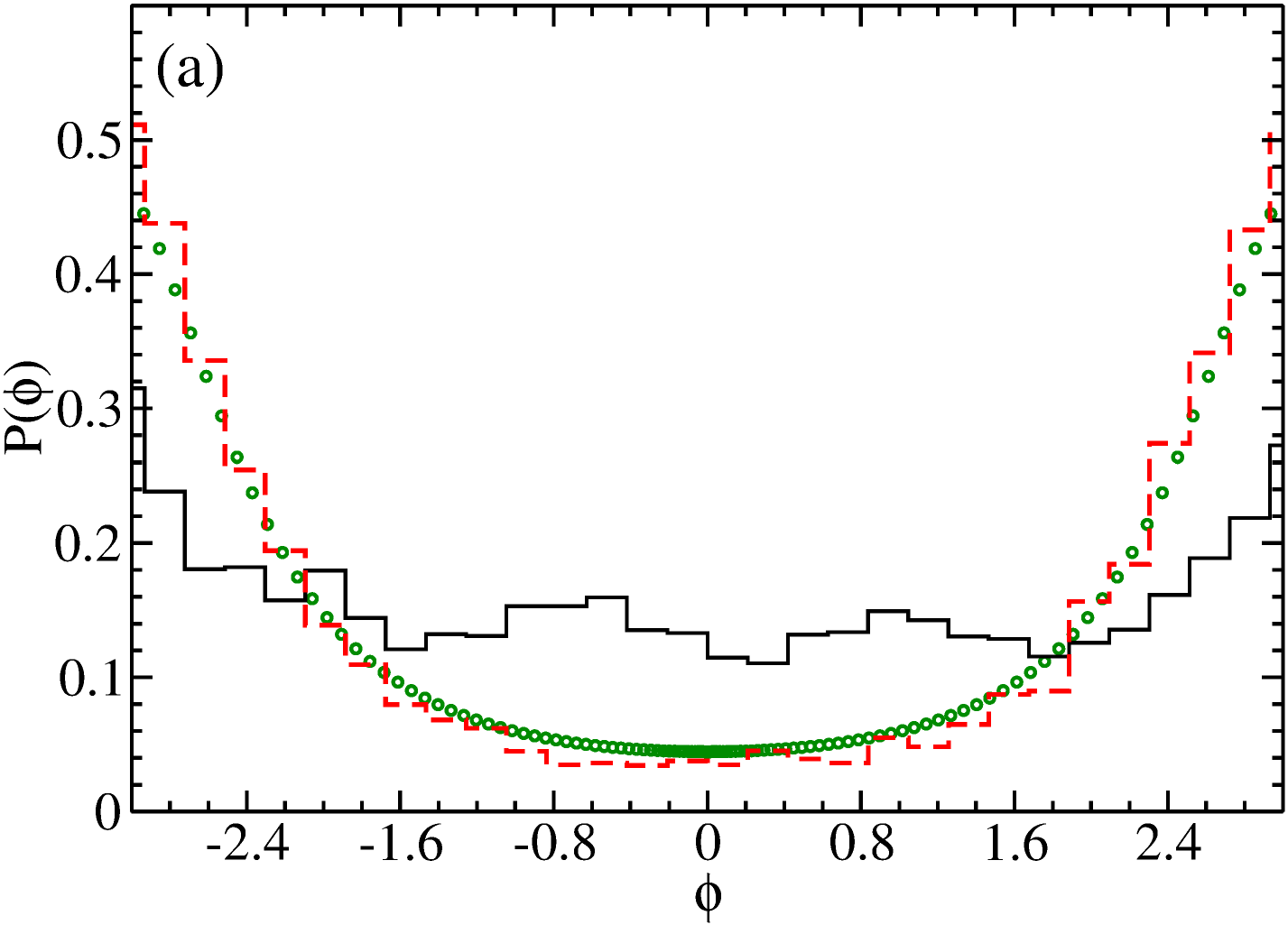}
\includegraphics[width=0.7\linewidth]{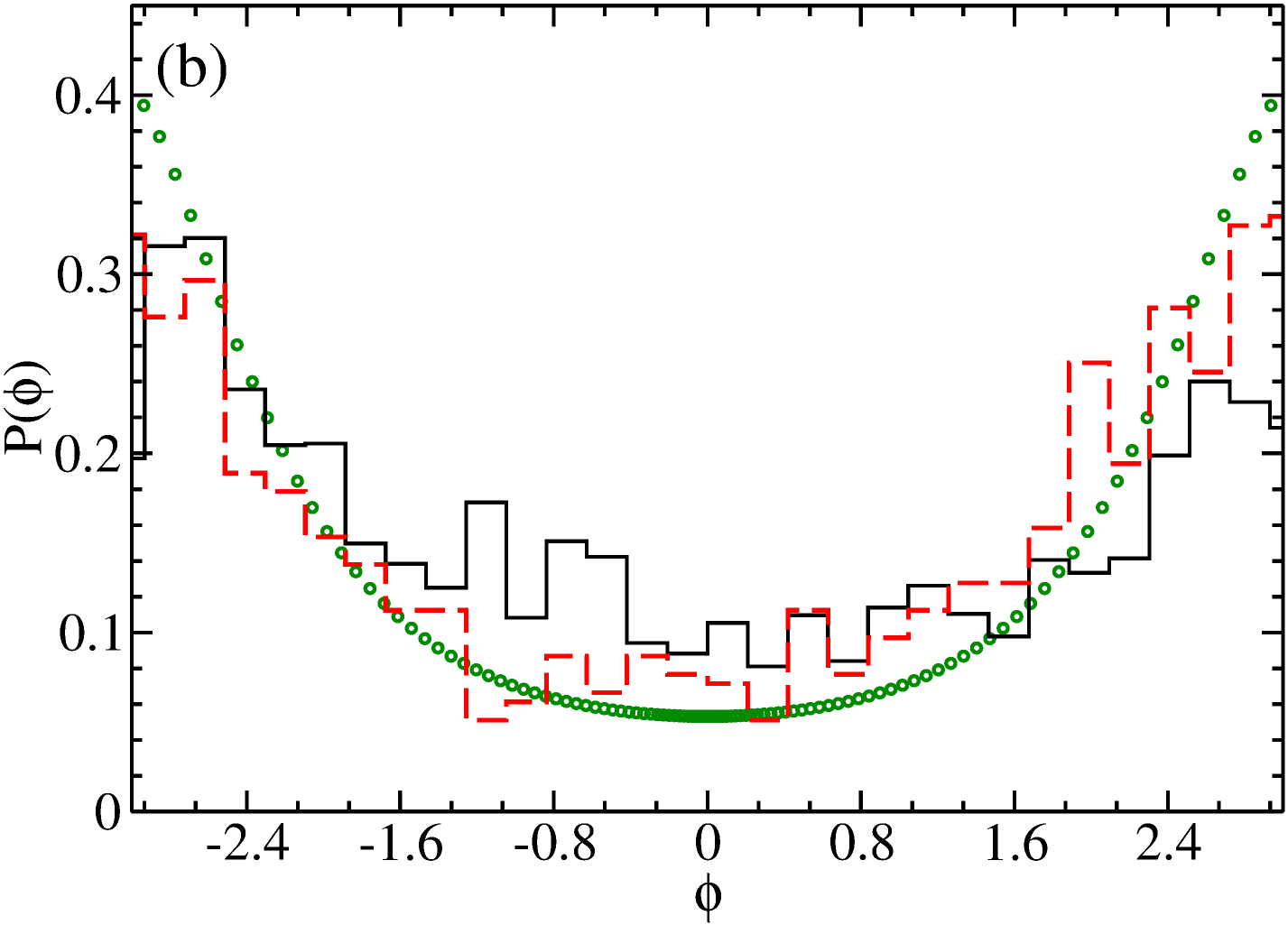}
	\caption{Same as ~\reffig{fig:corrport} for the distribution of the phases $\phi$ of the $S$-matrix elements, $S_{11}=\vert S_{11}\vert e^{i\phi}$.}
\label{fig:phs12disport}
\end{figure}

We, furthermore, analyzed the distributions of the amplitudes $\vert S_{aa}\vert$ and phases $\phi$ of the $S$ matrix $S_{aa}=\vert S_{aa}\vert e^{i\phi}$ and compared them to analytical results derived in~\cite{Fyodorov2005} and simplified in~\cite{Dietz2010} for the RMT $S$ matrix model and to those of the corresponding quantum graphs in Figs.~\ref{fig:s12disport} and~\ref{fig:phs12disport}. The agreement again is good for the honeycomb waveguide and both quantum graphs, whereas clear deviations are visible for the tetrahedral waveguide graph. 

Thus, according to our findings, the combination of a sufficiently large number of Y junctions and bent waveguides with bending angle $120^\circ$ induces sufficient complexity of the wave dynamics to achieve good agreement with the RMT results and, as predicted in~\cite{Pluhar2013,Pluhar2013a,Pluhar2014} for quantum graphs, the open honeycomb waveguide graph exhibits the properties of a typical quantum-chaotic scattering system.

\section{CONCLUSION\label{Concl}}
We report on experiments with superconducting waveguide networks. One consists of four $90^\circ$ bends, three T junctions and one Y junction and has the geometry of a tetrahedral quantum graph, whereas the waveguide graph with honeycomb structure consists exclusively of $120^\circ$ bends and Y junctions. We analyzed their spectral properties in the frequency range of a single transversal mode, where the Helmholtz equation is effectively one dimensional and equivalent to that of microwave cable networks and to the Schr\"odinger equation of quantum graphs. Yet, in distinction to quantum graphs with Neumann BCs at the vertices and microwave cable networks, the vertex scattering matrix describing the propagation of the microwaves through the junctions is frequency dependent. As outlined in~\refsec{WG} the bending angles and junctions of the honeycomb waveguide graphs were designed such that backscattering and frequency dependence are minimized and the scattering properties are the same at all openings of the junctions, as assumed in quantum graphs. Indeed, we find for it good agreement of the spectral properties with those of the corresponding quantum graph with Neumann BCs and with RMT. 

For the identification of the eigenfrequencies we used transmission and reflection spectra that were measured with 8 antennas attached to the otherwise closed waveguide graph. We furthermore performed scattering experiments. For this we opened the waveguide graphs at three bends and attached ports. Due to the high-quality factor attained with the superconducting waveguide graphs, the openness predominantly arises due to the ports, that are strongly coupled to the exterior. The fluctuation properties of the measured $S$ matrix were compared to those obtained from a RMT model based on the Heidelberg approach. We find very good agreement for the two-point correlation functions and the distributions of the $S$ matrix amplitudes for the honeycomb waveguide graph. 

A drawback of quantum graphs with Neumann BCs is the occurrence of backscattering at the vertices. For the tetrahedral waveguide graph, which comprises T joints and $90^\circ$ bends, the effect of backscattering becomes visible when comparing the spectral properties and the fluctuation properties of the $S$ matrix with those of the corresponding quantum graph. We demonstrated in numerical simulations~\cite{Bittner2013}, also with CST Studios, that these effects are minimized when using junctons that join waveguides with a relative angle of $120^\circ$  

Thus we conclude that superconducting waveguide graphs, comprising a large number of vertices and constructed from Y junctions, may serve as an experimental model for quantum graphs~\cite{Post2012,Exner2015,Gnutzmann2022} with the advantage compared to microwave cable networks that the eigenfrequencies and the $S$ matrix describing the scattering process of the corresponding open graph, that is, of microwaves coupled into the waveguide graph through ports at the open ends of waveguides, can be determined with high precision. The advantage of the normal conducting waveguide graphs used in Ref.~\cite{Zhang2022} is that for them also wave functions an be measured. Yet, we obtained information on properties of the wave functions from the resonance spectra in terms of the strength distribution, which accounts for the wave functions at the positions of the antennas or ports. 

\section{Acknowledgement}
This work has been supported through the Collaborative Research Centers SFB 634 and SFB 1245 of the German Science Foundation (DFG). We are deeply grateful to Hans Arwed Weidenmueller with whom we had fruitful discussions concerning the design of the vertex structure of the waveguide graphs, the associated boundary conditions and applicability of RMT. BD acknowledges financial support from the Institute for Basic Science in Korea through the project IBS-R024-D1.
\bibliography{References}
\end{document}